\documentclass[11pt]{amsart}
\usepackage{dirtytalk}
\usepackage{caption}
\usepackage{hyperref}
\usepackage{pdflscape}
\usepackage{float}
\usepackage{amsmath}
\usepackage{amssymb}
\usepackage{latexsym}
\usepackage{pictex}
\usepackage{pdfpages}
\usepackage{graphicx}
\usepackage[export]{adjustbox}
\title{The Quantitative Genetics of Human Disease:  1 Foundations}
\author{David J. Cutler $^{1,2,\ast}$}
\author{Kiana Jodeiry $^{2,3}$}
\author{Andrew J. Bass $^{1,2}$}
\author{Michael P. Epstein $^{1,2}$}
\address{$^1$Department of Human Genetics, Emory University}
\address{$^2$Center of Computational and Quantitative Genetics, Emory University}
\address{$^3$Department of Psychology, Emory University} 
\address{$^{\ast}$Corresponding author}
\begin{document}

\maketitle
\def\refname{Literature Cited}
\renewcommand{\thesubsection}{\thesection.\alph{subsection}}

\section{Abstract}

In this the first of an anticipated four paper series, fundamental results of quantitative genetics are presented from a first principles approach.  While none of these results are in any sense new, they are presented in extended detail to precisely distinguish between definition and assumption, with a further emphasis on distinguishing quantities from their usual approximations.  Terminology frequently encountered in the field of human genetic disease studies will be defined in terms of their quantitive genetics form.  Methods for estimation of both quantitative genetics and the related human genetics quantities will be demonstrated. While practitioners in the field of human quantitative disease studies may find this work pedantic in detail, the principle target audience for this work is trainees reasonably familiar with population genetics theory, but with less experience in its application to human disease studies. We introduce much of this formalism because in later papers in this series, we demonstrate that common areas of confusion in human disease studies can be resolved be appealing directly to these formal definitions.   The second paper in this series will discuss polygenic risk scores.  The third paper will concern the question of ``missing'' heritability and the role interactions may play.  The fourth paper will discuss sexually dimorphic disease and the potential role of the X chromosome.

\section{Introduction}

Arguably the most important paper in the history of population genetics theory was Fisher 1918, ``The Correlation between Relatives on the Supposition of Mendelian Inheritance.'' \cite{fisher1918} In this  work, nearly impenetrable to read by modern standards, Fisher established the fundamental model of quantitative genetics, unified the seemingly incompatible genetical models of Mendel and Galton, derived heritability from first principles, showed how to predict the correlation between relatives as a function of heritability, and began the process of defining and formalizing analysis of variance \cite{moran1966}.   All told, not a bad accomplishment for a work begun as an undergraduate that may have been in revision or ``review'' for the better part of 8 years \cite{FisherDaughter}.

Buried at the heart of Fisher's model is the idea of the effect of an allele on the phenotype of an individual.  In Fisher's presentation, and subsequent presentations by Falconer \cite{FalconerMacKay} and many others, the effect of an allele on phenotype is imagined as a physically determined entity - an allele with an effect two inches on height transmits two inches of height to an offspring when inherited from a parent.  The effect of the allele is in some sense immutable, independent of its context or how it is observed.  We can think of this interpretation of an allele as analogous to the classical mechanical interpretation of the atom.  An electron has energy, spin or position that is determined at all times. In Kemthorne's 1955 \cite{Kempthorne1955} derivation of fundamental quantitative genetics results, he introduces a subtly different interpretation of the effect of an allele.  In the Kempthorne presentation, the effect of an allele is fundamentally probabilisitic and only determined by the presence of other genetic and environmental effects.   Analogous to the Copenhagen interpretation of the atom where an electron's state is only determined when acted upon by external forces such as observation, the Kempthorne interpretation of allelic effect makes an allelic effect the average of some probability distribution, and this effect is only fully determined to be in a particular state in the presence of all other genetic and environmental factors.     In the sort of ``infinitesimal'' limit imagined by Fisher/Falconer, where the effects of individual alleles are so small as to be nearly unmeasurable, there is likely no practical difference between the Fisher/Falconer and Kempthorne interpretations of a genetic effect.  In the context of $21^{\text{st}}$ century human genetics, where the goal of an experiment is often to accurately measure the genetic effect of an allele, the distinction between the these two interpretations will be seen to lie at the crux of many of the most apparently confounding observations.

For all that follows in these series of papers, we will follow the Kempthorne interpretation of genetic effects.  We do so for several reasons.  First, in the opinion of these authors, Kempthorne's approach is, in some sense, more biologically realistic.  Almost everything in biology seems interconnected with other elements.  It seems more plausible that an allele has a slightly differing effect in every genetic/environmental context than it has the exact same effect in any two different contexts.   Second, the Kempthorne interpretation will help us to better understand numerous perplexing observations in human genetics.  Finally, we favor the Kempthorne interpretation for its modeling elegance and ease of presentation. 

\section{Quantitative Traits: Definitions and Foundational Results}
The presentation below largely follows Kempthrone, 1955, in a somewhat more modern notation, with much greater detail to assist the student in understanding results.  While the formalism is strictly Kempthorne's, in only a very few places does the distinction between the Kempthorne vs. Fisher/Falconer interpretation lead to any material difference in how a result is viewed or understood.  In the those cases we will endeavor to point out the effect the differing interpretations has.   Throughout this section we will refer to the Fisher/Falconer interpretation of genetic effect as the Falconer interpretation as his detailed derivations, presentations, and formalism are far more commonly read by population geneticists than Fisher's.  In our first simplification from Kempthorne, we restrict our presentation to only two alleles at each locus because in a modern context we think of these loci as single nucleotide changes, single nucleotide polymorphisms (SNP) in the usual term of human genetics, rather than a more abstract concept like gene or locus that Kempthorne envisioned nearly $70$ years ago.
\subsection{Single locus}
To begin, consider a single diploid locus in Hardy-Weinberg equilibrium with two alleles $A_0$ and $A_1$, where the frequency of $A_0$ is $p$, and the frequency of $A_1$ is $q = 1-p$.   For the sake of notational convenience let us suppose that we have oriented the allelic labels such that $p \ge q$.  Thus, in the parlance of human genetics, $A_0$ is the ``major'' allele, and $A_1$ is the ``minor'' allele.  Imagine individuals have some observable, measurable quantitative phenotype $Y$ such as height, weight, or blood pressure.   Further suppose that individuals with genotype $A_0A_0$ have average phenotype $y_{00}$, individuals with genotype $A_0A_1$ have average phenotype $y_{01}$, and individuals with genotype $A_1A_1$ have average phenotype $y_{11}$.  Thus,   
\begin{eqnarray*}
\mathrm{E}[Y | G=A_0A_0] & = & y_{00}. \\
\mathrm{E}[Y | G=A_0A_1] & = &y_{01}. \\
\mathrm{E}[Y | G=A_1A_1] & = & y_{11}. \\
\mu_y = \mathrm{E}[Y] & = &\mathrm{Pr}[G=A_0A_0]\mathrm{E}[Y | G=A_0A_0] + \mathrm{Pr}[G=A_0A_1]\mathrm{E}[Y | G=A_0A_1] \\
& & + \mathrm{Pr}[G=A_1A_1]\mathrm{E}[Y | G=A_1A_1]  \\
& = & p^2y_{00} + 2pqy_{01} + q^2y_{11}.
\end{eqnarray*}
\noindent  The overall population mean is thus found by appeal to the law of total expectation: the expectation of random variable $X$ is the $\sum \mathrm{Pr}[X=x] \mathrm{E}[X | X = x]$, where the sum is taken over all possible states $x$ of the random variable $X$.  For computational tractability, instead of working with phenotype $Y$, we will instead consider the linear transformation of $Y$, $P$, where $P = Y - \mu_y$.  Thus, $P$ is a zero centered transformation of $Y$,  $\mathrm{E}[P] = \mathrm{E}[Y - \mu_y] = \mathrm{E}[Y] - \mu_y = 0$, but the shape of $P$'s distribution is the same as $Y$'s.

Define the ``genetic effects'' $\gamma_{00},\gamma_{01},\gamma_{11}$ of genotypes $A_0A_0,A_0A_1, A_1A_1$ to be the average phenotype of individuals with those genotypes.   If we let $G$ be the two allele genotype at this locus
\begin{eqnarray*}
\gamma_{00} & = & \mathrm{E}[P | G = A_0A_0].  \\
\gamma_{01} & = & \mathrm{E}[P | G = A_0A_1]. \\
\gamma_{11} & = & \mathrm{E}[P | G = A_1A_1].  \\
\mathrm{E}[P] & = &\mathrm{Pr}[G=A_0A_0]\mathrm{E}[P | G=A_0A_0] + \mathrm{Pr}[G=A_0A_1]\mathrm{E}[P | G=A_0A_1]  \\
&& +  \mathrm{Pr}[G=A_1A_1]\mathrm{E}[P | G=A_1A_1]  \\
 & = & p^2\gamma_{00} + 2pq\gamma_{01} + q^2\gamma_{11} = 0.
\end{eqnarray*}  
\noindent Thus, the genetic effect of genotype $G = A_iA_j$, $i,j \in \{0,1\}$, is given by $\gamma_{ij}$, which is the conditional expectation of phenotype, given the individual has genotype $A_iA_j$.  Notice that if two populations have differing genotype frequencies at this locus, the genetic effects are \emph{necessarily} different, since both populations will have been normalized to have mean zero phenotype.   Here we see the first element of the difference between the Falconer and Kempthorne interpretations.  A Falconer view point might imagine the genetic effects as fixed and independent of allele frequencies.  In Kempthorne's approach genetic effects are only defined conditional on the genotype frequencies.  

In a similar fashion, call the ``allelic effect'' the conditional expectation of phenotype, given an individual possesses the allele.   Let $\alpha_0$ and $\alpha_1$ be the allelic effects of $A_0$ and $A_1$.  To find $\alpha_0$ imagine picking an individual at random from the population. Next imagine picking an allele at random from the chosen person.  The probability that the chosen allele was $A_0$ is, by definition, $p$.  Similarly, the probability the picked allele was $A_1$ is $q$.  We find the allelic effect $\alpha$ as the conditional expectation of phenotype given the picked allele.
\begin{eqnarray*}
\alpha_0 & = & \mathrm{E}[P | A_0\text{ picked}] \\
& = & \mathrm{Pr}[G=A_0A_0 |  A_0\text{ picked}] \mathrm{E}[P | G = A_0A_0] +  \mathrm{Pr}[G=A_0A_1 |  A_0\text{ picked}] \mathrm{E}[P | G=A_0A_1]  \\
&   &  +   \mathrm{Pr}[G =A_1A_1 |  A_0\text{ picked}] \mathrm{E}[P | G=A_1A_1] \\
& = & {p^2 \over p} \gamma_{00} + {1 \over 2 } {2pq \over p} \gamma_{01} + 0 \\
& = & p \gamma_{00} + q \gamma_{01}. \\
\alpha_1 & = &  \mathrm{E}[P | A_1\text{ picked}] \\
& = & p \gamma_{01} + q \gamma_{11}.
\end{eqnarray*}
\noindent Importantly, note that from these definitions 
\begin{eqnarray*}
p \alpha_0 + q \alpha_1 & = & p(p\gamma_{00} + q\gamma_{01}) + q(p\gamma_{01} + q \gamma_{11}) \\
& = & p^2 \gamma_{00} + 2pq \gamma_{01} + q^2\gamma_{11} = 0. \\
\alpha_0 & = & {- q \alpha_1 \over p}.\\
\alpha_1  & = &  { -p \alpha_0 \over q},
\end{eqnarray*}
\noindent further reenforcing the notion that in the Kempthorne framework the allelic effects are defined in terms of the allele frequencies.   We define a related variable $\beta = \alpha_1 - \alpha_0$ as the difference in the allelic effects between the two alleles.  This variable $\beta$ is naturally interpreted as the consequence of substituting an $A_1$ allele for an $A_0$ allele, and will be commonly estimated in a linear regression.

While formally we define $\alpha$ as an allelic effect (mean phenotype of an individual with that allele), we will often refer to $\alpha$'s as the ``additive effect'' of an allele, and may frequently use the terms ``allelic effect'' and ``additive effect'' interchangeably.   At first blush this interchange of terms may seem very odd.   Traditionally in one locus population genetics the term ``additive'' is used to describe a dominance relationship.   A locus is called additive when the phenotype of the heterozygote is the average of the two homozygous phenotypes.   In this context, a locus is additive if $\gamma_{01} = { \gamma_{00} + \gamma_{11} \over 2 }$.   It turns out that there is a very natural reason to equate the terms ``allelic effects'' and ``additive effects.''  Note that if the locus is additive then 
\begin{eqnarray*}
0 & = &   p^2\gamma_{00} + 2pq\gamma_{01} + q^2\gamma_{11} \\
& = & p^2\gamma_{00} + 2pq(\gamma_{00}+\gamma_{11})/2 + q^2\gamma_{11} \\
& = & \gamma_{00}(p^2 + pq) + \gamma_{11}(pq+q^2) \\
& = & \gamma_{00}p(p+q) + \gamma_{11}q(p+q) \\
& = & p\gamma_{00} + q\gamma_{11} \\
2\alpha_0  & = & 2p \gamma_{00} + 2q \gamma_{01} \\
& = & 2p\gamma_{00} + 2q (\gamma_{00}+\gamma_{11})/2 \\
& = & p\gamma_{00} + q\gamma_{00} + p\gamma_{00} + q\gamma_{11} \\
& = & \gamma_{00} (p+q) + 0 \\
&= & \gamma_{00} \\
\alpha_0+\alpha_1 & = & p \gamma_{00} + q \gamma_{01} + p\gamma_{01} + q\gamma_{11} \\
& = & p\gamma_{00} + \gamma_{01}(p+q) + q\gamma_{11} \\
& = & (p\gamma_{00} + q\gamma_{11}) + \gamma_{01} \\
& = & \gamma_{01} \\
2\alpha_1  & = & 2p \gamma_{01} + 2q \gamma_{11} \\
& = & 2p (\gamma_{00}+\gamma_{11})/2 + 2q\gamma_{11} \\
& = & p\gamma_{00} + p\gamma_{11} + q\gamma_{11}  + q\gamma_{11}\\
& = & (p\gamma_{00} + q\gamma_{11}) + \gamma_{11}(p+q) \\ 
&= &  0 + \gamma_{11} = \gamma_{11}. \\
\end{eqnarray*}
\noindent Thus, we find for an additive locus the total genetic effects are simply the sum of the individual allele effects added together.    For such an additive locus
\begin{eqnarray*}
\gamma_{00} & = & 2\alpha_0. \\
\gamma_{01} & = & \alpha_0 + \alpha_1. \\
\gamma_{11} & = & 2\alpha_1.
\end{eqnarray*}

In a Falconer inspired presentation of this work, one might have been asked to assume that the total genetic effect at a locus was the sum of the individual ``additive'' effects of the alleles.   This could be an assumption of the model.   In a Kempthrone framework, where the definition of allelic effects are the mean phenotype of individuals with that allele, for any locus in Hardy-Weinberg that is additive, additivity \emph{implies} that the genotype effect is the sum of the allelic effects.  For an additive locus, the genotype effect is simply the sum of its individual allelic effects.  For a non-additive locus, the genotypic effects will differ from the sum of the allelic effects.  Let $\delta$ be the difference between the genetic effects of a genotype from the sum of its individual allelic effects.   In particular, let
\begin{eqnarray*}
\delta_{00} & = & \gamma_{00} - 2\alpha_0. \\
\gamma_{00} & = & 2\alpha_0 + \delta_{00}. \\
\delta_{01} & = & \gamma_{01} - (\alpha_0 + \alpha_1). \\
\gamma_{01} & = & \alpha_0  + \alpha_1 + \delta_{01}. \\
\delta_{11} & = & \gamma_{11} - 2\alpha_1. \\ 
\gamma_{11} & = & 2\alpha_1 + \delta_{11}. 
\end{eqnarray*}

Now, let us imagine a random variable, $g$, representing the genetic effect of this locus, where its value is determined by the genotype of an individual.  Thus if an individual has genotype $G = A_iA_j$, then $g = \gamma_{ij}$.  Genotype is viewed as a randomizing process, and when $G = A_iA_j$ a random variable $g$ has value $\gamma_{ij}$.   This random variable $g$ can be further decomposed into a random variable $a$, whose value is the sum of the allelic effects $a = \alpha_i + \alpha_j$, and another random variable $d = \delta_{ij}$,  the deviation from additivity due to dominance.   In all cases we think of these random variables, $g$,$a$,$d$, as being determined by the random process of genotype in the individual.  Thus, in a notational convention we will attempt to maintain throughout, $G$ refers to a randomly determined genotype with average phenotype $\gamma$.  $A$ refers to a random allele, with average phenotype $\alpha$.  The lower case $g$, $a$ and $d$ are random variables determined by the random genotype giving rise to this locus's genetic, additive, and dominance contributions. The fact that $P$ has mean $0$ implies the average component contributions from this locus must also be $0$.  
\begin{eqnarray*}
\mathrm{E}[g] & = &  \mathrm{E}[ \mathrm{E}[ P | G ]] =  \mathrm{E}[P] = 0 \\
\mathrm{E}[a] & = & \mathrm{E}[\mathrm{E}[a | G = A_iA_j]] = \mathrm{E}[\alpha_i + \alpha_j] = \mathrm{E}[ \alpha_i] + \mathrm{E}[\alpha_j]   \\
& = &  \mathrm{Pr}[A_i=A_0]\mathrm{E}[P |A_i=A_0] + \mathrm{Pr}[A_i=A_1]\mathrm{E}[P |A_i=A_1]   \\
&& +  \mathrm{Pr}[A_j=A_0]\mathrm{E}[P |A_j=A_0] + \mathrm{Pr}[A_j=A_1]\mathrm{E}[P |A_j=A_1] \\
& = & p \alpha_0 + q\alpha_1 + p\alpha_0 + q\alpha_1 = 0 \\
\mathrm{E}[d] & = & \mathrm{E}[g - a] = \mathrm{E}[g] - \mathrm{E}[a]  = 0 
\end{eqnarray*} 
\noindent While the average genetic, additive and dominance effects are all zero, they each might contribute to total phenotypic variance.   In particular the genetic variance due to this locus, $V_g$ is
\begin{eqnarray*}
V_g & = & \mathrm{Var}[g] = \mathrm{E}[g^2 ] - (\mathrm{E}[g])^2 = \mathrm{E}[ g^2] \\
& = &  \mathrm{Pr}[G=A_0A_0]\mathrm{E}[P |G=A_0A_0]^2 + \mathrm{Pr}[G=A_0A_1]\mathrm{E}[P |G=A_0A_1]^2 \\
&& + \mathrm{Pr}[G=A_1A_1]\mathrm{E}[P |G=A_1A_1]^2 \\
& = & p^2(\gamma_{00})^2 + 2pq(\gamma_{01})^2 + q^2(\gamma_{11})^2.
\end{eqnarray*}
\noindent The additive variance, $V_a$, due to this locus is 
\begin{eqnarray*}
V_a & = & \mathrm{Var}[a] =  \mathrm{E}[a^2] -(\mathrm{E}[a])^2 =   \mathrm{E}[a^2] \\
& = &  \mathrm{Pr}[G=A_0A_0] (2\alpha_0)^2 +  \mathrm{Pr}[G=A_0A_1] (\alpha_0+\alpha_1)^2 + \mathrm{Pr}[G=A_1A_1] (2\alpha_1)^2 \\
& = & p^2(4\alpha_0^2) + 2pq(\alpha_0^2 + 2\alpha_0\alpha_1 + \alpha_1^2) + q^2(4\alpha_1^2) \\
& = & 2p\alpha_0(2p\alpha_0 + q\alpha_0 + q\alpha_1) + 2q\alpha_1(2q\alpha_1 + p\alpha_1 + p\alpha_0) \\  
& = & 2p\alpha_0(\alpha_0(p+q) + p\alpha_0 + q\alpha_1) + 2q\alpha_1(\alpha_1(p+q) + p\alpha_0 + q\alpha_1) \\
& = & 2(p\alpha_0^2 + q\alpha_1^2) 
\end{eqnarray*}
\noindent Notice the $2$ in front of the sum.  Intuitively the quantity inside the parenthesis is the additive variance due to a single allele, and the $2$ comes from the fact that this is a diploid organism with additive contributions from both alleles.  The dominance variance, $V_d$ from this locus is
\begin{eqnarray*}
V_d & = & \mathrm{Var}[d] = \mathrm{E}[d^2] - (\mathrm{E}[d])^2  =  \mathrm{E}[d^2] \\
& = & p^2(\delta_{00})^2 + 2pq(\delta_{01})^2 + q^2(\delta_{11})^2. 
\end{eqnarray*}

In a result that might be considered something less than completely obvious, $\mathrm{Var}[g] = \mathrm{Var}[a] + \mathrm{Var}[d]$.  This follows from the definition $g = a + d$.   Necessarily $\mathrm{Var}[g] = \mathrm{Var}[a + d] = \mathrm{Var}[a] + \mathrm{Var}[d] + 2 \mathrm{Cov}[a,d]$, but 
\begin{eqnarray*}
\mathrm{Cov}[a,d] & =&  \mathrm{E}[ad] - \mathrm{E}[a]\mathrm{E}[d] = \mathrm{E}[ad] \\
& = & p^2(2\alpha_0\delta_{00}) + 2pq((\alpha_0+\alpha_1)d_{01}) + q^2(2\alpha_1d_{11}) \\
& = & p^2(2\alpha_0(\gamma_{00}-2\alpha_0) + 2pq((\alpha_0+\alpha_1)(\gamma_{01}-(\alpha_0+\alpha_1))) + q^2(2\alpha_1(\gamma_{11}-2\alpha_1))  \\
&= & p^2(2\alpha_0\gamma_{00}) + 2pq((\alpha_0+\alpha_1) \gamma_{01}) + q^2(2\alpha_1\gamma_{11}) - \left[p^2(2\alpha_0)^2 + 2pq(\alpha_0+\alpha_1)^2 + q^2(2\alpha_1)^2\right] \\
& = & 2p\alpha_0(p\gamma_{00} + q\gamma_{01}) + 2q\alpha_1(p\gamma_{01} + q\gamma_{11}) - V_a \\
& = & 2p\alpha_0^2 + 2q\alpha_1^2 - V_a = 0.
\end{eqnarray*}
\noindent Thus, the additive and dominance contributions to variance are fundamentally orthogonal within a locus in Hardy-Weinberg equilibrium.  The total genetic variance is simply the sum the additive and dominance variance contributions.   Put another way, if a locus is in Hardy-Weinberg Equilibrium then there is no interaction between additivity and dominance, or perhaps even more intuitively, within a single locus in Hardy-Weinberg, the only possible deviation from additivity is an uncorrelated dominance effect.   On the other hand, inbreeding and other departures from Hardy-Weinberg create correlation between the allelic states and can create correlation between the additive and dominance components within a locus.

\subsection{Many Loci and Environments}

Moving to multiple loci we expand our notation as follows.   Let $G_v$ be the genotype at locus $v$.  Again assuming two alleles $A_{v_0}$ and $A_{v_1}$ at every genetic locus, $v$, $\gamma_{v_{00}}$,  $\gamma_{v_{01}}$, and $\gamma_{v_{11}}$ corresponds to the genotypic effects of the three genotypes at this locus.   Let the allelic effects at this locus be $\alpha_{v_0}$ and $\alpha_{v_1}$.   Let $g_v$, $a_v$ and $d_v$ be the random variables induced by the genotype at locus $v$ with values determined by the corresponding values of $\gamma$, $\alpha$ and $\delta$, reflecting the genetic, additive and dominance contributions of this locus.  Call the corresponding variance terms $V_{g_v}$, for the total genetic variance, $V_{a_v}$ for the additive variance, and $V_{d_v}$ for the dominance variance.  See Table 1 for a summary of several key variables introduced in this section [TABLE HERE].  All these individual locus effects are defined in the previous section.  To work our way to many loci, we start by building from two loci, $v$ and $w$.   To begin, consider the notion of a two-locus genotypic effect, which for loci $v$ and $w$, we will call $\gamma_{v_{ij},w_{kl}}$ when the two loci genotypes are $G_v = A_{v_i}A_{v_j}$, $i,j \in \{0,1\}$, and $G_w = A_{v_k}A_{w_l}$, $k,l \in \{0,1\}$ 
\begin{equation*}
\gamma_{v_{ij},w_{kl}} = \mathrm{E}[P | G_v = A_{v_i}A_{v_j}, G_w = A_{w_k}A_{w_l}. ]
\end{equation*}
\noindent Here the $\gamma$ tells us it is a genetic effect (mean phenotype given genotype).  The subscript $v_{ij}$ tells us one of loci involved is locus $v$ and the genotype of locus $v$ is $G_v = A_{v_i}A_{v_j}$.  After the comma we find a second locus is given, $w$, where the genotype of $w$ is $G_w = A_{w_k}A_{w_l}$.  Putting this all together we read $\gamma_{v_{ij},w_{kl}}$ as the expected phenotype of an individual given their genotype is $A_{v_i}A_{v_j}$ at locus $v$ and  $A_{w_k}A_{w_l}$ at locus $w$. In general we will use $v$ and $w$ to correspond to distinct loci.   All loci have two alleles, and for these two loci we will use $i,j \in \{0,1\}$ to correspond particular alleles $A_{v_0}$ and $A_{v_1}$ at locus $v$, and $k,l \in \{0,1\}$ for the alleles at locus $w$.  Think of the random variable $g_{v,w}$ corresponding the the two locus genetic effect $\gamma$ determined by the random genotype at the two loci, such that $g_{v,w} = \gamma_{v_{ij},w_{kl}}$ when the genotype of $G_v$ is $A_{v_i}A_{v_j}$ and the genotype of $G_w$ is $A_{w_k}A_{w_l}$,
\begin{eqnarray*}
\mathrm{E}[g_{v,w}] & = & \mathrm{E}[ \mathrm{E}[ P | G_v = A_{v_i}A_{v_j}, G_w = A_{w_k}A_{w_l} ] ] = \mathrm{E}[P] = 0 \\
\mathrm{Var}[g_{v,w}] & = & \mathrm{E}[ (g_{v,w})^2 ] - (\mathrm{E}[g_{v,w}])^2 = \mathrm{E}[ (g_{v,w})^2 ] \\
& = & \sum_{i,j,k,l} \mathrm{Pr}[G_v = A_{v_i}A_{v_j},G_w = A_{w_k}A_{w_l}]   (\gamma_{v_{ij},w_{kl}})^2 
\end{eqnarray*}
The next question is ``how does the two locus genetic effect relate to the individual loci effects?'' Let us first assume that the way in which locus $v$ and $w$ interact to create phenotype is their joint genetic effect is the sum of the individual genetic effects.  In other words, one possible way these loci might interact is in an additive fashion, such that
\begin{equation*}
\gamma_{v_{ij},w_{kl}} = \gamma_{v_{ij}} + \gamma_{w_{kl}}
\end{equation*}
\noindent Call this manner of interaction, ``additive'', because the joint genetic effect is just the sum of the individual genetic effects.    Of course, the loci need not interact in an additive fashion.  Quantitative geneticists traditionally use the term epistatic to mean any sort of non-additive interaction between loci, but this term has a less well-defined meaning in the human genetics community.  For the sake of convenience we will call these interactions between loci either additive, or non-additive.    Analogous to the dominance deviation within a single locus, let us think of a multilocus quantity that we will call the ``interaction deviation,'' or others might call the ``epistatic deviation,'' which will measure the deviation from additivity of the multilocus genotype.  In particular, define the interaction deviation $\delta_{Ig_{v_{ij},w_{kl}}}$ between these loci as 
\begin{equation*}
\delta_{Ig_{v_{ij},w_{kl}}}  =  \gamma_{v_{ij},w_{kl}} - (\gamma_{v_{ij}} + \gamma_{w_{kl}}).
\end{equation*}
\noindent Corresponding to this interaction deviation, we will think of a random variable $d_{Ig_{v,w}}$ whose value is given by $\delta_{Ig_{v_{ij},w_{kl}}}$ whenever the two loci have genotypes $A_{v_i}A_{v_j}$ and $A_{w_k}A_{w_l}$.
\begin{eqnarray*}
\mathrm{E}[d_{Ig_{v,w}}] & = & \mathrm{E}[ g_{v,w} - (g_v+ g_w) ] = 0 - (0+0) = 0. \\
\mathrm{Var}[d_{Ig_{v,w}} ] & = & \mathrm{E}[ (g_{v,w} - (g_v+ g_w))^2 ] - (\mathrm{E}[ (g_{v,w} - (g_v+ g_w) ])^2 \\
& = & \sum_{i,j,k,l} \mathrm{Pr}[G_v = A_{v_i}A_{v_j},G_w = A_{w_k}A_{w_l}] (\delta_{Ig_{v_{ij},w_{kl}}})^2.
\end{eqnarray*}
\noindent We can decompose the entire two locus genetic variance into its component variances.
\begin{eqnarray*}
\mathrm{Var}[g_{v,w}] & = & \mathrm{Var}[g_v + g_w + d_{Ig_{v,w}} ] \\
& = & \mathrm{Var}[g_v] + \mathrm{Var}[g_w] + (\mathrm{Var}[d_{Ig_{v,w}}] + 2(\mathrm{Cov}[g_v,g_w] + \mathrm{Cov}[g_v,d_{Ig_{v,w}}] + \mathrm{Cov}[g_w,d_{Ig_{v,w}}])) \\
& = & V_{g_v} + V_{g_w} +   (\mathrm{Var}[d_{Ig_{v,w}}] + 2(\mathrm{Cov}[g_v,g_w] + \mathrm{Cov}[g_v,d_{Ig_{v,w}}] + \mathrm{Cov}[g_w,d_{Ig_{v,w}}])).
\end{eqnarray*} 
\noindent In this fashion we define the interaction ``variance'' between locus $v$ and $w$, $V_{Ig_{v,w}}$, to be
\begin{equation*}
V_{Ig_{v,w}} = \mathrm{Var}[d_{Ig_{v,w}}] + 2(\mathrm{Cov}[g_v,g_w] + \mathrm{Cov}[g_v,d_{Ig_{v,w}}] + \mathrm{Cov}[g_w,d_{Ig_{v,w}}])
\end{equation*}
We will define $V_{Ig_{v,w}}$ as the interaction ``variance'', but the term ``variance'' should remain in quotes. When we examined the within locus additive by dominance covariance we found these were necessarily $0$.  That is \emph{not} necessarily true for locus $\times$ locus interactions.  As a result this entity that we are calling a ``variance'' is not a variance.  It is the sum of a variance, and three covariances, and as a result it can, and frequently will, be negative!  A particularly important case, the non-random association of alleles due to proximity of the loci on a chromosome, linkage disequilibrium (LD), will often have the effect of leading to negative interaction variance.  This will be discussed in much greater detail below.  For all that follows we will often drop the quotes from variance, but the reader should never lose sight of the fact that this is not a proper variance, but, in fact, a variance/covariance sum and as a result need not be positive.  If LD is going to be treated explicitly in any estimation of genetic effects, it is in this stage where it might be most precisely handled.  

A reasonable reader might object to the use of the term ``variance'' to ever describe this sum.  Such an objection is well grounded.   We use the term ``variance'' for historical reasons.   Nearly every other derivation of quantitative genetics from Fisher/Falconer through to Kempthorne explicitly or implicitly assumes the state of one genetic locus (or environment, see below) is independently chosen from any other.  With that explicit assumption in mind, the interaction variance, $V_{Ig_{v,w}}$, is simply the squared sum of the interaction deviation $ \mathrm{Var}[d_{Ig_{v,w}}]$ and is a proper variance.  So, the historical use of that term is correct and well warranted.   However, beginning with the assumption of uncorrelated genotypic or environmental states makes accounting for their correlation when it does actually exist a considerable challenge.  At some level it makes one wonder to what extent quantities are even well defined when the first assumption of the modeling framework is violated.   Here, we make no assumption about state correlation, and therefore have an ability to explicitly model that correlation (as we may for LD), and we see the manner in which state correlation effects the total variance is by adding interaction covariances.  Thus, if one were to estimate the interaction variance by subtracting the main effects from the total variance, in the presence of state correlation, the interaction variance can be negative.  On the other hand, if we chose to follow Kempthorne's suggestion and estimate the interaction variance as the squared interaction deviation, $\mathrm{Var}[d_{Ig_{v,w}}]$, the estimated quantity will always be non-negative, but the component variances will certainly not sum to the total variance unless genotypic/environmental states are uncorrelated.    Since we do not wish to assume away the very real existence of LD, and wish to use terms that at least roughly correspond their historical usage, we find ourselves with an interaction variance, that is not a variance and might be negative.   Had this field developed after, say, the discovery of the structure of the \emph{lac} operon \cite{jacobmonod1961}, we might find ourselves with less confusingly defined terms.

Setting these nomenclature objections aside, we can further decompose that genetic interaction variance into its additive and dominance components.  To do so we will consider a series of deviations from the average phenotype given some combination of alleles and genotypes at the two loci, and its expected value if all interactions were additive.    Using the notation $A_v$ to indicate a randomly picked allele at locus $v$, we define the deviations as
\begin{eqnarray*}
\delta_{Iaa_{v_i,w_k}} & = & \mathrm{E}[P | A_v = A_{v_i},A_w=A_{w_k}] - (\alpha_{v_i}  + \alpha_{w_k}) \\
& = & { \sum_{j,l}\mathrm{Pr}[G_v=A_{v_i}A_{v_j},G_w=A_{w_k}A_{w_l}]\gamma_{v_{ij},w_{kl}} \over \sum_{j,l}\mathrm{Pr}[G_v=A_{v_i}A_{v_j},G_w=A_{w_k}A_{w_l}] } - (\alpha_{v_i} + \alpha_{w_k}). \\
\delta_{Iad_{v_i,w_{kl}}} & = & \mathrm{E}[P | A_v = A_{v_i},G_w=A_{w_k}A_{w_l}] - (\alpha_{v_i} + \alpha_{w_k} + \alpha_{w_l} + \delta_{w_{kl}} + \delta_{Iaa_{v_i,w_k}} + \delta_{Iaa_{v_i,w_l}}  )\\
& = &  { \sum_{j}(\mathrm{Pr}[G_v=A_{v_i}A_{v_j},G_w=A_{w_k}A_{w_l}]\gamma_{v_{ij},w_{kl}} \over \sum_{j}(\mathrm{Pr}[G_v=A_{v_i}A_{v_j},G_w=A_{w_k}A_{w_l}] } \\
&& - (\alpha_{v_i} + \alpha_{w_k} + \alpha_{w_l} + \delta_{w_{kl}} + \delta_{Iaa_{v_i,w_k}} + \delta_{Iaa_{v_i,w_l}}  ) ). \\
\delta_{Ida_{v_{ij},2_{k}}} & = & \mathrm{E}[P | G_v = A_{v_i}A_{v_j},A_2=A_{w_k}] - (\alpha_{v_i} + \alpha_{v_j} + \alpha_{w_k} + \delta_{v_{ij}} + \delta_{Iaa_{v_i,w_k}} + \delta_{Iaa_{v_j,w_k}}  )\\
& = &  { \sum_{l}(\mathrm{Pr}[G_v=A_{v_i}A_{v_j}G_w=A_{w_k}A_{w_l}]\gamma_{v_{ij},w_{kl}} \over   \sum_{l}(\mathrm{Pr}[G_v=A_{v_i}A_{v_j}G_w=A_{w_k}A_{w_l}] }\\
&& - (\alpha_{v_i} + \alpha_{v_j} + \alpha_{w_k} + \delta_{v_{ij}} + \delta_{Iaa_{v_i,w_k}} + \delta_{Iaa_{v_j,w_k}}  )\\
\delta_{Idd_{v_{ij},w_{kl}}} & = & \mathrm{E}[P | G_v = A_{v_i}A_{v_j},G_w=A_{w_k}A_{w_l}] - (\alpha_{v_i} + \alpha_{v_j}  + \alpha_{w_k} + \alpha_{w_l} + \delta_{v_{ij}} + \delta_{w_{kl}} \\
&&  +   \delta_{Iaa_{v_i,w_k}} + \delta_{Iaa_{v_i,w_l}} + \delta_{Iaa_{v_j,w_k}}  + \delta_{Iaa_{v_j,w_l}}+ \delta_{Iad_{v_i,w_{kl}}} + \delta_{Iad_{v_j,w_{kl}}} + \delta_{Ida_{v_{ij},2_{k}}}  +  \delta_{Ida_{v_{ij},2_{l}}}   )\\
& = & \gamma_{v_{ij},w_{kl}}- (\alpha_{v_i} + \alpha_{v_j}  + \alpha_{w_k} + \alpha_{w_l} + \delta_{v_{ij}} + \delta_{w_{kl}}  \\
&& +   \delta_{Iaa_{v_i,w_k}} + \delta_{Iaa_{v_i,w_l}} + \delta_{Iaa_{v_j,w_k}}  + \delta_{Iaa_{v_j,w_l}}+ \delta_{Iad_{v_i,w_{kl}}} + \delta_{Iad_{v_j,w_{kl}}} + \delta_{Ida_{v_{ij},w_k}}  +  \delta_{Ida_{v_{ij},w_l}}   ).
 \end{eqnarray*}
\noindent We can therefore write all $9$ two locus genotype effects as a sum of the expected effects assuming additivity and the appropriate $9$ deviations from additivity.
\begin{eqnarray*}
\gamma_{v_{ij},w_{kl}} & = & \alpha_{v_i} + \alpha_{v_j} +  \delta_{v_{ij}}  + \alpha_{w_k} + \alpha_{w_l} + \delta_{w_{kl}} \\
& & \delta_{Iaa_{v_i,w_k}} + \delta_{Iaa_{v_j,w_k}} + \delta_{Iaa_{v_i,w_l}} + \delta_{Iaa_{v_j,w_l}}  \\
& & + \delta_{Iad_{v_i,w_{kl}}} +  \delta_{Iad_{v_j,w_{kl}}}  + \delta_{Ida_{v_{ij},w_k}} + \delta_{Ida_{v_{ij},w_l}}  + \delta_{Idd_{v_{ij},w_{jk}}}.
\end{eqnarray*}
\noindent Corresponding to each of these deviations we think of random variables $d_{Iaa_{v,w}}$,  $d_{Iad_{v,w}}$, and $d_{Idd_{v,w}}$ induced by the random genotypes $G_v$ and $G_w$, and we simplify notation by combining ``like'' terms to get
\begin{eqnarray*}
d_{Iaa_{v,w}} & = &  \delta_{Iaa_{v_i,w_k}} + \delta_{Iaa_{v_j,w_k}} + \delta_{Iaa_{v_i,w_l}} + \delta_{Iaa_{v_j,w_l}}. \\
d_{Iad_{v,w}} & = &  \delta_{Iad_{v_i,w_{kl}}} +  \delta_{Iad_{v_j,w_{kl}}}  + \delta_{Ida_{v_{ij},w_k}} + \delta_{Ida_{v_{ij},w_l}}.  \\
d_{Idd_{v,w}} & = & \delta_{Idd_{v_{ij},w_{jk}}}.
\end{eqnarray*}
\noindent Arriving at the full decomposition of the two locus genetic effects viewed as random variables,
\begin{eqnarray*}
g_{v,w} & = & a_v + d_v + a_w + d_w + d_{Iaa_{v,w}} + d_{Iad_{v,w}}  + d_{Idd_{v,w}} \\
V_{g_{v,w}} & = & \mathrm{Var}[a_v] + \mathrm{Var}[d_v] +  \mathrm{Var}[a_w] + \mathrm{Var}[d_w] + V_{Iaa_{v,w}} + V_{Iad_{v,w}} + V_{Idd_{v,w}} \\ 
V_{Iaa_{v,w}}  & = & \mathrm{E}[(d_{Iaa_{v,w}})^2] + \mathrm{Cov}[ d_{Iaa_{v,w}},d_{Iad_{v,w}}] + \mathrm{Cov}[ d_{Iaa_{v,w}}, d_{Idd_{w,w}}]  \\ 
V_{Iad_{v,w}} & = & \mathrm{E}[(d_{Iad_{v,w}})^2] + \mathrm{Cov}[ d_{Iad_{v,w}},d_{Iaa_{v,w}}] + \mathrm{Cov}[ d_{Iad_{v,w}}, d_{Idd_{v,w}} ] \\
& & + 2(\mathrm{Cov}[d_{Iaa_{v,w}},a_v] + \mathrm{Cov}[d_{Iaa_{v,w}},a_w] + \mathrm{Cov}[d_{Iaa_{v,w}},d_v] + \mathrm{Cov}[d_{Iaa_{v,w}},d_w] ) \\
& & + 2(\mathrm{Cov}[d_{Iad_{v,w}},a_v] + \mathrm{Cov}[d_{Iad_{v,w}},a_w] + \mathrm{Cov}[d_{lad_{v,w}},d_v] + \mathrm{Cov}[d_{Iad_{v,w}},d_w]) \\
V_{Idd_{v,w} } & = & \mathrm{E}[(d_{Idd_{v,w}})^2] + \mathrm{Cov}[ d_{Idd_{v,w}},d_{Iaa_{v,w}}] + \mathrm{Cov}[ d_{Idd_{v,w}}, d_{Iad_{w,w}}]  \\
& & + 2(\mathrm{Cov}[d_{Idd_{v,w}},a_v]  + \mathrm{Cov}[d_{Idd_{v,w}},a_w] + \mathrm{Cov}[d_{Idd_{v,w}},d_v] + \mathrm{Cov}[d_{Idd_{v,w}},d_w] ) 
\end{eqnarray*}
\noindent Thus, we have constructed the additive by additive, $V_{Iaa}$, additive by dominance, $V_{Iad}$, and dominance by dominance, $V_{Idd}$, interaction variances as the sum of a deviation variance and several covariance terms, which means none of these terms are true ``variances'', and in the presence of LD might be negative.  In practice we will often assume that all these covariances are absent or negligible, and estimate each term as the squared deviation, or even as the residual variance after subtracting the lower order terms.   Extending this framework to arbitrarily large numbers of loci is essentially more of the same.   If there are a total of $N$ loci, we construct total variance terms as 
\begin{eqnarray*}
V_G & = & \sum_{v=1}^N V_{g_v}. \\
V_A & = & \sum_{v=1}^N V_{a_v}. \\
V_D & = & \sum_{v=1}^N V_{d_v}. \\
V_{GG} & = &  \sum_{v=1}^N \sum_{w=v+1}^N V_{Ig_{v,w}}. \\
V_{AA} & = & \sum_{v=1}^N \sum_{w=v+1}^N V_{Iaa_{v,w}}. \\
V_{AD} & = &  \sum_{v=1}^N \sum_{w=v+1}^N V_{Iad_{v,w}}. \\
V_{DD} & = &  \sum_{v=1}^N \sum_{w=v+1}^N V_{Idd_{v,w}}.  \\
V_{GGG} & = & \sum_{v=1}^N \sum_{w=v+1}^N  \sum_{z=w+1}^N V_{Ig_{v,w,z}}.\\
.... \\
V_{DD...D} & = & V_{Idd_{1,2,...,N}},
\end{eqnarray*}
\noindent where each of the newly introduced interaction terms are defined with reference to the difference between the mean phenotype given that combination of genotypes and/or alleles, and the expectation if all those factors interacted in a strictly additive fashion plus all the lower order interaction deviations.  Of course, all of those interaction variances are not true variances but the sum of a deviation variance and a number of covariances, making them all potentially negative in the presence of LD. 

Next we extend this framework to include ``environmental'' influences on phenotype.   In the usual parlance of quantitative genetics, an environmental factor is anything that can effect the phenotype that is not genetic.  Aspects of diet, exposure to the elements, contact with a virus, stochastic ``noise'' in the statistical sense, or an enormous number of other things could all be environmental influences on phenotype.  With this broad definition in mind, we imagine $M$ distinguishable environmental factors $E_m$, $1 \le m \le M$.  By assumption environmental factor $m$ can take on more than one state, and we will write $E_m =x$ to indicate that environmental factor $m$ is in state $x$.   Analogous to genetic effects we talk about the main effects $\epsilon_{m_x}$ of being in state $x$ for environmental factor $m$, and the corresponding random variable $e_m$.
\begin{eqnarray*}
\epsilon_{m_x} & = & \mathrm{E}[P | E_m = x] \\
\mathrm{E}[e_m] & = & \mathrm{E}[\epsilon_{m_x}] =  \mathrm{E}[\mathrm{E}[P | E_m = x] ] = 0 \\
V_{e_m} & = & \mathrm{Var}[\mathrm{E}[P | E_m = x] ] = \mathrm{E}[(\mathrm{E}[P | E_m = x])^2].
\end{eqnarray*} 
These environmental factors interact with each other in some fashion.  This interaction could be the sum of their individual main effects (additive) or deviate from additivity.  We therefore consider the combined effects of environmental factors $m$ and $s$, $e_{m_x,s_y}$ and the deviation from additivity between these factors.
\begin{eqnarray*}
\epsilon_{m_x,s_y} & = & \mathrm{E}[P | E_m = x, E_{s} = y] \\
\mathrm{E}[e_{m,s}] & = &  \mathrm{E}[\mathrm{E}[P | E_m = x,E_{s}=y] ] = 0 \\
\delta_{Ie_{m_x,s_y}}& = & \epsilon_{m_x,s_y} - (\epsilon_{m_x} + \epsilon_{s_y}) \\
V_{Ie_{m,s}} & = & \mathrm{E}[(d_{Ie_{m,s}})^2] + 2(\mathrm{Cov}[e_m,e_{s}] + \mathrm{Cov}[e_m,d_{Ie_{m,s}}] + \mathrm{Cov}[e_{s},d_{Ie_{m,s}}]),
\end{eqnarray*} 
\noindent where $d_{Ie_{m,s}}$ is the random variable whose value is $\delta_{Ie_{m_x,s_y}}$ when environment $m$ is in state $x$ and environment $s$ is in state $y$.   If these environmental states are uncorrelated with one another then $V_{Ie_{m,s}} = \mathrm{E}[(d_{Ie_{m,s}})^2]$ , but if the state of environment $m$ correlates with the state of $s$, then the covariances might be substantial leading, potentially, to negative interaction ``variance.''

Genetic and environmental factors interact.  This interaction might be purely additive, or include some deviation from additivity.  For locus $v$ with alleles $A_{v,i}$ and $A_{v,j}$ and environmental factor $m$ with state $x$,
\begin{eqnarray*}
\gamma\epsilon_{v_{ij},m_x} & = & \mathrm{E}[P | G_v=A_{v,i}A_{v,j},E_m = x] \\
\mathrm{E}[ge_{v,m} ] & = & \mathrm{E}[\gamma\epsilon_{v_{ij},m_x}] =   \mathrm{E}[\mathrm{E}[P | G_n =A_{v,i},A_{v,j},E_m = x]] = 0 \\
\delta_{Ige_{v_{ij},m_x}} & = & \gamma\epsilon_{v_{ij},m_x} - ( \gamma_{v_{ij}} + \epsilon_{m_x}) \\
d_{Ige_{v,m}} & = &  ge_{v,m} - (g_v + e_m) \\
V_{Ige_{v,m}} & = &  \mathrm{E}[ (d_{Ige_{v,m}})^2] + 2 (\mathrm{Cov}[ d_{Ige_{v,m}},g_v ] + \mathrm{Cov}[ d_{Ige_{v,m}},e_m ] + \mathrm{Cov}[g_v,e_m ] ),
\end{eqnarray*} 
where $ge_{v,m}$ and $d_{Ige_{v,m}}$ are the random variables associated with $\gamma\epsilon_{v_{ij},m_x}$ and  $\delta_{Ige_{v_{ij},m_x}}$, respectively, and are determined by the random states of locus $v$ and environment $m$.  Additive by environment and dominance by environment interactions can be constructed in a similar fashion.   Higher order variances are constructed with the same logic, as the deviation between the conditional phenotype and its expectation assuming additive interaction squared, plus twice the appropriate covariances, giving rise to 
\begin{eqnarray*}
V_E & = & \sum_{m=1}^M V_{e_m} \\
V_{GE} & = &  \sum_{v=1}^N \sum_{m=1}^M V_{Ige_{n,m}} \\
V_{GGE} & = & \sum_{v=1}^N \sum_{w=n+1}^N  \sum_{m=1}^M V_{Igge_{v,w,m}}\\
...&  & \\ 
\end{eqnarray*}
\noindent It should go without repeating that all of these interaction ``variances'' are not true variances, and in the presence of correlation between genes and the environment could be negative.  

\subsection{The resemblance between Relatives}

Notice that up to this point we have made very few assumptions about individual genetic, environmental or interaction effects.  We have implicitly assumed that the number of genetic and environmental factors is countable.  This assumption is certain for genetic factors which, for man, is surely bounded in some fashion by the number of nucleotides in the genome $\approx 3\times10^9$.  Implicitly we have also assumed that all the discussed quantities are finite, and therefore have finite variances, but this is a very weak assumption indeed, considering all terms are ultimately defined in terms of conditional expectations of phenotypes of ``real'' organisms, and in the case we are most interested in here, actual human beings.  As a result we have made no particular assumptions about the distribution of phenotypes in the population, other than no values are infinite.   

Choose an individual, $p$, at random from the population.  Call their phenotype $P_p$.   $\mathrm{E}[P_p] = 0$.   Call the variance in their phenotype $\mathrm{Var}[P_p] = V_P$, the total phenotypic variance.  This individual has some genotype $G_v$ at all $N$ loci, and experienced some set of environmental influences, $E_m$ for all $M$ environments.   Thus,
\begin{eqnarray*}
\mathrm{E}[P_p] & = & \mathrm{E}[\mathrm{E}[P_p |  G_1 = A_{1,i}A_{1,j}...,G_N = A_{N,i}A_{N,j},E_1 = e_{1x},...E_M = e_{m_x}]  ] \\
& = & \mathrm{E}[ g_0 + g_1 + ... g_n + e_1 + ... + e_M  + d_{Ig_{0,1}} + ... + d_{Ig_{N-1,N}} + d_{Igg_{1,2,3} } + ...\\
& &  + d_{Ige_{1,1}} + ... + d_{Ie_{1,2}} + .... + d_{Iee....e_{1,2,....,M}} ] \\
& = & \mathrm{E}[ a_1 + d_ 1  + ... +  a_N + d_N + e_1 + ... + e_M + d_{Iaa_{1,2}} + ... + d_{Iaa_{N-1,N}} +  d_{Iad_{1,2}}+ ... \\
& & +d_{Iae_{1,1}} +... d_{Ie_{1,2}} + .... + d_{Iee...e_{1,2,..,M}} ] = 0 \\
\mathrm{Var}[P_p] & = & V_A + V_D + V_E + V_{AA} + V_{AD} + V_{AE} + V_{DE} + V_{AAD} + V_{ADD} + ... + V_{EE...E}.  
\end{eqnarray*}

Now imagine two individuals $1$ and $2$ with phenotype $P_1$ and $P_2$.   These two individuals might be unrelated, in which case they are both random draws from the population and $\mathrm{Cov}[P_1,P_2] = 0$.    For individuals who are related, a convenient way to quantify their degree of relatedness is with something that human geneticist call Cotterman coefficients \cite{Cotterman1940} but here we will follow a more \cite{Wright1922} inspired presentation.  At any given genetic locus, individuals $p_1$ and $p_2$ might share $0$, $1$ or $2$ alleles that are identical by descent (IBD), a term used to mean that the alleles are identical \emph{because} the alleles were inherited by both individuals without modification from a recent common ancestor.   Let $\rho_0$ be the probability that $0$ alleles were inherited IBD at some locus.  Let $\rho_1$ be the probability that exactly one allele was inherited IBD, and $\rho_2$ be the probability that both alleles were inherited IBD.   By assumption these probabilities are the same at all autosomal loci in the genome.  Let $\rho = \rho_2 + {\rho_1 \over 2}$ be the``coefficient of relatedness'' between these two individuals.   The simplest interpretation of $\rho$ is the expected fraction of the autosomal genome shared IBD between the individuals.    To find the resemblance between these relatives, which we will quantify as the $\mathrm{Cov}[P_1,P_2]$, we begin with a single genetic locus and single environmental effect.
\begin{eqnarray*}
\mathrm{Cov}[P_1,P_2] & = & \mathrm{E}[P_1P_2] - \mathrm{E}[P_1]\mathrm{E}[P_2] \\
& =  & \mathrm{E}[P_1P_2] \\
& = & \mathrm{E}[(a_{p1} + d_{p1} +  e_{p1} + d_{Iae_{{1,1}_{p1}}} + d_{Ide_{{1,1}_{p1}}} )(a_{p2} + d_{p2} + e_{p2} + d_{Iae_{{1,1}_{p2}}} + d_{Ide_{{1,1}_{p2}}}) ] \\
& = & \mathrm{E}[ a_{p1}  a_{p2}] + \mathrm{E}[ d_{p_1} d_{p2}] +  \mathrm{E}[ e_{p_1}  e_{p2}]  \\
& &  + \mathrm{E}[ d_{Iae_{{1,1}_{p1}}} , a_{p2}] + ... + \mathrm{E}[d_{Ide_{{1,1}_{p1}}}, d_{Ide_{{1,1}_{p2}}} ]. 
\end{eqnarray*}
\noindent The last step used the fact that $\mathrm{E}[a,d]$ within a locus is $0$.  If these two individuals experience the environment independently of one another the only non-zero terms above are  $\mathrm{E}[ a_{p1}  a_{p2}]$ and $\mathrm{E}[ d_{p1} d_{p2}]$.  Even if the individuals have correlated environments, if there is no correlation between an individual's genes and the environments they experience, the only other non-zero term is $\mathrm{E}[ e_{p1}  e_{p2}]$.   If we assume environments are independent of genotype, then this can be simplified to
\begin{eqnarray*}
\mathrm{Cov}[P_1,P_2] & = & \mathrm{E}[ a_{p1}  a_{p2}] + \mathrm{E}[ d_{p1} d_{p2}]  \\
& = & \mathrm{Pr}[\text{IBD}_0] (\mathrm{E}[ a_{p1}  a_{p2} | \text{IBD}_0] ] + \mathrm{E}[ d_{p1}  d_{p2} | \text{IBD}_0] ] \\
& & + \mathrm{Pr}[\text{IBD}_1] (\mathrm{E}[ a_{p1}  a_{p2} | \text{IBD}_1] ] + \mathrm{E}[ d_{p1}  d_{p2} | \text{IBD}_1] ] \\
& & + \mathrm{Pr}[\text{IBD}_2] (\mathrm{E}[ a_{p1}  a_{p2} | \text{IBD}_2] ] + \mathrm{E}[ d_{p1}  d_{p2} | \text{IBD}_2] ] \\
& = & \rho_0 (0 + 0) + \rho_1 ({V_a \over 2} + 0) + \rho_2 (V_a + V_d) \\
& = & \rho V_a + \rho_2 V_d.
\end{eqnarray*}
\noindent We leave as an exercise for the student to show the transition between the second and third lines above is correct, but the result is perfectly intuitive.  If two individuals share exactly one allele IBD, then they share half the additive variance at this locus.  If they share two alleles IBD then they share all the additive variance and all the dominance variance.  Otherwise, there is no expected correlation between the individuals.  Extension of this result to multiple loci, again with the assumption of uncorrelated environments between the individuals, proceeds in a similar fashion to reach the well known \cite{Kempthorne1955}
\begin{equation}
\mathrm{Cov}[P_1,P_2] = \rho V_A + \rho_2 V_D + \rho^2 V_{AA} + \rho(\rho_2)V_{AD} + (\rho_2)^2 V_{DD}  + \rho^3 V_{AAA} + \rho^2(\rho_2)V_{AAD} +...+ ... .
\end{equation}
\noindent The $\rho^2$ before the $V_{AA}$ term comes from the fact that in order to share an interaction between two loci the individuals must share one or more alleles at both loci.  The $\rho(\rho_2)$ before $V_{AD}$ derives from the requirement of sharing at least one allele at one locus, and two at the other, and so forth.  Notice that we have arrived at the fundamental result of Fisher 1918/Kempthorne 1955 without making any distributional assumptions at all about phenotype or the size or nature of genetic and environmental effects.  This result holds if these quantities exist and are finite.  Thus, the observation that most phenotypes are approximately normally distributed is not an assumption of quantitative genetics, but \emph{evidence} that there are likely many genetic and/or environmental factors contributing to any nearly normally distributed phenotype, and many of those factors are interacting in a nearly additive fashion.  Normality is a consequence of various Feller like versions \cite{Feller1946} of the strong law of large numbers which establishes that as the number of random variables included in a sum grows large, if a sufficiently large subset of those factors are uncorrelated, the sum will converge to a normal distribution.  Thus, from our prospective when a phenotype is observed to be normally distributed, or nearly so, this should be taken as an indication that the phenotype is likely contributed to by many genetic and/or environmental factors interacting in an often additive fashion.

For known familial relationships, such as parent, $P_p$ and offspring, $P_o$, we immediately reach the well known
\begin{eqnarray*}
\mathrm{Cov}[P_p,P_o] & = & {V_A \over 2}  +  {V_{AA} \over 4}  +  {V_{AAA}  \over 8}  + ... + ... \\
& \approx & {V_A \over 2}. 
\end{eqnarray*}
\noindent The last line being the form of this result most commonly taught to students.   Viewed in this fashion, the student taught result is not so much an assumption about a lack of interaction variance, but a consequence of the fact that interactions ``transmit'' from parent to offspring diminished by a factor of ${1 \over 2}$ for each successive level of interaction.  So, unless the interaction variances are of the same order of magnitude as the main effect, dropping these higher order interactions is a natural approximation that will hold under most circumstances.  Similarly for full siblings $s1$ and $s2$ we have
\begin{eqnarray*}
\mathrm{Cov}[P_{s1},P_{s2}]  & = &{V_A \over 2} + {V_D \over 4}  + {V_{AA} \over 4}  + {V_{AD} \over 8}  + {V_{DD}  \over 16}  + { V_{AAA} \over 8}  +....+ ...\\
& \approx &  {V_A \over 2} + {V_D \over 4} \\ 
& \approx &  {V_A \over 2},
\end{eqnarray*}
\noindent with the last approximation assuming that dominance is weak in comparison to additive effects.  

For historical and practical reasons involved in animal husbandry, quantitative geneticists created a particular abstraction often called the ``mid-parent'' which is the mean phenotype of the two parents of some offspring.  Thus if $P_{p1}$ and $P_{p2}$ are the phenotypes of the two parents then $P_{mid} = {P_{p1} + P_{p2} \over 2}$, and if $P_o$ is the phenotype of their offspring it is trivial to show that 
\begin{eqnarray*}
\mathrm{Var}[P_{mid}] & = & {V_p \over 2} \\
\mathrm{Cov}[P_{mid},P_o] & = & {V_A \over 2} + {V_{AA} \over 4} + {V_{AAA} \over 8} +...+ \\
& \approx &  {V_A \over 2}. 
\end{eqnarray*}
\noindent All of this holds regardless of the distribution of phenotype or genetic/environmental effects.   However, for many bivariate distributions of random variables $X,Y$, including bivariate normal distributions, it is straightforward to show that 
\begin{equation*}
\mathrm{E}[X | Y ] = {Y \mathrm{Cov}[X,Y] \over \mathrm{Var}[Y]} 
\end{equation*}
\noindent So, if we assume this relationship holds for the distribution of phenotypes considered here (because the distribution is approximately normal, say) then we arrive at the definition of heritability $h^2$ and its natural interpretation
\begin{eqnarray*}
\mathrm{E}[P_o | P_{mid}] & = & {P_m \mathrm{Cov}[P_o,P_{mid}] \over \mathrm{Var}[P_{mid}]} \\
& = & { P_{mid}({V_A \over 2} + {V_{AA} \over 4} + {V_{AAA} \over 8} +...+) \over {V_P \over 2} } \\
& \approx & {P_{mid} V_A \over V_P } \\
h^2 & = & {V_A \over V_P } .
\end{eqnarray*}
Thus, heritability $h^2$ is interpretable as the factor that predicts average offspring phenotype as a function of average parental phenotype, and is the fraction of the total phenotypic variance that is due to the additive affects of alleles.  From here we get the interpretation that $V_A$, the additive variance, is the fraction of the phenotype transmitted from parent to offspring.  Or put slightly differently, parents transmit only their additive variance to their offspring.   We should of course note that this intuition was formed with an approximation which dropped all the higher order additive interactions.  

For any arbitrary pair of relatives $r1$ and $r2$ 
\begin{eqnarray*}
\mathrm{E}[P_{r2} | P_{r1} ] & = & {P_{r1} \mathrm{Cov}[P_{r2},P_{r1}] \over \mathrm{Var}[P_{r1}]} \\ 
& = & { P_{r1} (\rho V_A + \rho_2 V_D + \rho^2 V_{AA} + \rho(\rho_2)V_{AD} + (\rho_2)^2 V_{DD}  + ... + ...) \over V_P } \\
& \approx & P_{r1} \rho h^2 
\end{eqnarray*}
These results give rise to the most natural way to estimate $h^2$.   Collect a number of pairs of individuals with known familial relationship, pairs of a single parent and their offspring, say.  Measure the average phenotype of the parents, and average of the offspring.  The ratio of the offspring mean to the single parent mean is $\rho h^2= {h^2 \over 2}$.   Slightly more formally, regress offspring values on their parental values, and the slope of the regression is $h^2 \over 2$.   When the regression is performed offspring on mid-parent, the slope is $h^2 = {V_A \over V_P} \le 1$.  It is because offspring means are less than mid-parental means that the best fit line was named the``regression line''  It was the line that represented the fact that offspring had ``regressed'' towards the mean relative to their parents.  That this regression was the consequence of transmission of only additive factors was the major genetic insight of Fisher 1918 \cite{fisher1918}   Before his derivation, the existence of regression to the mean led to some very unusual ideas about how inheritance of complex phenotypes might work \cite{Galton18886}.  These ideas appear to those of us born after 1918 to be almost bizarre and certainly very hard to fathom once the phenomenon is correctly understood. 
  
\subsection{Accounting for Linkage Disequilibrium}

In a formal sense, within the Kempthorne modeling framework, linkage disequilibrium (LD) -the non-random association of variants at different loci, often induced by small physical distances between them on the same chromosome- can alter the size of the genetic effect, alter the distribution between additive and dominance sub components of that effect, and induce interaction ``variance'' between the loci, which in many biologically common cases creates a negative interaction ``variance.''   We will give some suggestions for explicitly modeling of this, but the intuition for why this occurs is important and also perfectly easy to see.   Imagine two loci $G_v$ and $G_w$ in what is called ``complete LD.''  If two loci are in complete LD, the genotype of every individual at locus $v$ is identical to the genotype at locus $w$.  Thus, $g_{v,w} = g_v = g_w$ in all individuals, and $V_{g_{v,w}} = V_{g_v} = V_{g_w}$.   The interaction deviance $d_{Ig_{v,w}} = g_{v,w} - (g_v + g_w) = -g_v$, and we immediately arrive at  $V_{g_v} = V_{g_w} = \mathrm{Var}[d_{Ig_{v,w}}]  = \mathrm{Cov}[g_v,g_w] = -\mathrm{Cov}[g_v,d_{Ig_{v,w}}] = -\mathrm{Cov}[g_v,d_{Ig_{v,w}}]$, and the interaction ``variance" is $V_{Ig_{v,w}} = -V_{g_v}$.  Thus, complete LD creates a negative interaction variance of the same size as the main effects.   As a general rule of thumb, LD causes neighboring sites to have more similar genetic effects then they would absent LD and induces negative interaction ``variance.''  

To begin to develop a framework for explicit accounting for LD, we start with some formal definitions.   Imagine two genetic loci $G_v$ and $G_w$ with alleles $A_{v_0},A_{v_1}$ and $A_{w_0},A_{w_1}$ respectively.   Let us further assume that these two genetic loci reside on the same chromosome.  Thus, there are four possible haploid entities that population geneticists often call ``gametes,'' and human geneticists ``haplotypes,'' that represent the identity of all possible allelic combinations at these two loci on a single piece of DNA.  Let $p_v$ and $p_v$ be the frequency of the $A_{v_0}$ allele and $A_{w_0}$ allele respectively.  Let $q_v = 1- p_v, q_w = 1- p_w$ be the frequency of the other allele at each locus.   Let $p_{00}, p_{01}, p_{10}, p_{11}$ be the frequencies of a haplotypes containing the $A_{v_0}A_{w_0}, A_{v_0}A_{w_1}, A_{v_1}A_{w_0}, A_{v_1}A_{w_1}$ alleles respectively (FIGURE 1).

The population geneticist defines, $D$, the standard measure of linkage disequilibrium \cite{Gillespie2004}, and the related $r^2$ as 
\begin{eqnarray*}
D &= & p_{00} - p_vp_w \\
& = & -p_{10} + q_vp_w \\
& = & -p_{01} + p_vq_w \\
& = & p_{11} - q_vq_w \\
p_v & = & p_{00} + p_{01}. \\
q_v & = & p_{10} + p_{11}. \\
p_w & = & p_{00} + p_{10}. \\
q_w & = & p_{01} + p_{11}. \\
p_{00} & = & p_vp_w + D \\
p_{01} & = & p_vq_w - D \\
p_{10} & = & q_vp_w - D \\
p_{11} & = & q_vq_w + D \\
r^2 & = & {D^2 \over p_vq_vp_wq_w} 
\end{eqnarray*}
\noindent While this historical definition has its applications, a far more intuitively informative presentation begins by thinking of the alleles at $G_v$ and $G_w$ as Bernoulli random variables on $\{0,1\}$ with the state of Bernoulli variable determined by the state of the allele at the locus on a given haplotype.   Thus, consider jointly distributed Bernoulli random variables $S_v,S_w \in \{0,1\}$ to correspond to the state of the alleles at $G_v$ and $G_w$ on some randomly picked haplotype.  With this in mind, 
\begin{eqnarray*}
\mathrm{E}[S_v] & = &  \mathrm{Pr}[\text{Picked }A_{v_0}] \times 0 +  \mathrm{Pr}[\text{Picked }A_{v_1}] \times 1 \\
& = & q_v. \\
\mathrm{E}[S_w] & = & q_w. \\
\mathrm{Var}[S_v] & = & \mathrm{E}[S_v^2] - (\mathrm{E}[S_v])^2 \\
& = & \mathrm{Pr}[\text{Picked }A_{v_1}] \times 1^2 - q_v^2 \\
& = & q_v - q_v^2 = p_vq_v. \\
\mathrm{Var}[S_w] & = & p_wq_w. \\
\mathrm{Cov}[S_v,S_w] & = & \mathrm{E}[S_vS_w] - \mathrm{E}[S_v]\mathrm{E}[S_w] \\
& = &  \mathrm{Pr}[\text{Picked }A_{v_1}A_{w_1}] \times 1 - q_vq_w \\
& = & p_{11} - q_vq_w \\
& = & D. \\
r^2 & = & {(Cov[S_v,S_w])^2 \over \mathrm{Var}[S_v]\mathrm{Var}[S_w]}.
\end{eqnarray*}
\noindent Thus, the classical population genetics measure of LD, $D$, is nothing more than what might be called the haplotypic covariance, and the LD measure $r^2$ is the squared correlation coefficient between the alleles at the two loci.   Higher order LD can be expressed in terms of higher order covariance terms.  

To form an intuition for how this effects quantitative genetics quantities, let us assume there is no dominance at either locus, and that the only interaction between these two loci is induced by LD.   Thus, let us begin by generalizing our notion of $\alpha$, the average phenotype of an individual with a randomly picked allele, to $\eta$ the average phenotype of an individual given a randomly picked haplotype.  Letting $H$ donate a haplotype randomly picked from an individual in this population,
\begin{eqnarray*}
\eta_{v_i,w_k} & = & \mathrm{E}[ P | H = A_{v_i}A_{w_k} ] \\  
& = &\sum_{jl}p_{jl} \gamma_{v_{ij},w_{kl}}.  \\
\end{eqnarray*}

If we assume there are no interactions between these loci other than that which is induced by LD, then $\eta_{v_1,w_1} - \eta_{v_1,w_0} = \eta_{v_0,w_1} - \eta_{v_0,w_0}$ and $\eta_{v_1,w_0} - \eta_{v_0,w_0} = \eta_{v_1,w_1} - \eta_{v_0,w_1}$, In other words, the lack of interaction other than LD implies the difference in average phenotype between the alleles at the second locus are unaffected by the state of the first locus, and vice versa.    If a randomly picked individual has phenotype $P$ with genotype $G_v = A_{v_i}A_{v_j}G_w = A_{w_k}A_{w_l}$ with corresponding haplotypes $A_{v_i}A_{w_k}$ and $A_{v_j}A_{w_l}$ then

\begin{eqnarray*}
\alpha_{v_0} & =  & {p_{00} \eta_{v_0,w_0} + p_{01} \eta_{v_0,w_1} \over p_v}. \\
\alpha_{w_1} & = & {p_{10} \eta_{v_1,w_0} + p_{11} \eta_{w_1,w_1} \over q_v }. \\
\beta_v & = & \alpha_{v_1} - \alpha_{v_1} \\
& = & {p_v p_{10}  \eta_{v_1,w_0} - q_v p_{00} \eta_{v_0,w_0}  + p_vp_{11}   \eta_{v_1,w_1} - q_v p_{01} \eta_{v_0,w_1} \over p_vq_v}\\
& = & { p_v(q_vp_w-D)  \eta_{v_1,w_0}  - q_1(p_vp_w+D) \eta_{v_0,w_0} + p_v(q_vq_w+D)  \eta_{v_1,w_1} - q_v (p_vq_w-D )\eta_{v_0,w_1} \over p_vq_v} \\
& = & \left[p_w( \eta_{v_1,w_0}  - \eta_{v_0,w_0} ) + q_w(  \eta_{v_1,w_1} - \eta_{v_0,w_1} \right)] +  \left({D \over p_vq_v}\right)  \left[ p_v(  \eta_{v_1,w_1}  - \eta_{v_1,w_0}) +q_v( \eta_{v_0,w_1} - \eta_{v_0,w_0})  \right] \\
& = & \left[p_w( \eta_{v_1,w_0}  - \eta_{v_0,w_0} ) + q_w(  \eta_{v_1,w_1} - \eta_{v_0,w_1} \right)] +  \left({D \over p_vq_v}\right)  \left[ p_v(  \eta_{v_0,w_1}  - \eta_{v_0,w_0}) +q_v( \eta_{v_1,w_1} - \eta_{v_1,w_0})  \right] \\
\alpha_{w_0} & = & { p_{00} \eta_{v_0,w_0} + p_{10} \eta_{v_1,w_0} \over  p_w} \\
\alpha_{w_1} & = & { p_{01} \eta_{v_0,w_1} + p_{11} \eta_{v_1,w_1}  \over q_w} \\
\beta_w & = & \alpha_{w_1} - \alpha_{w_0} \\
& = &{ p_wp_{01} \eta_{v_0,w_1} - q_w p_{00} \eta_{v_0,w_0}  + p_2p_{11}  \eta_{v_1,w_1} - q_2 p_{10} \eta_{v_1,w_0} \over p_wq_w}.  \\
& = &{ p_w(p_vq_w-D)  \eta_{v_0,w_1} - q_w(p_vp_w+D) \eta_{v_0,w_0}  + p_w(q_vq_w+D)  \eta_{v_1,w_1} - q_w (q_vp_w-D) \eta_{v_1,w_0} \over p_wq_w}. \\ 
& = & \left[p_v (\eta_{v_0,w_1} -  \eta_{v_0,w_0}) + q_v(\eta_{v_1,w_1}  -  \eta_{v_1,w_0})  \right] +  \left({D \over p_wq_w}\right) \left[ p_w(  \eta_{v_1,w_1}  -   \eta_{v_0,w_1})  + q_w(  \eta_{v_1,w_0} -\eta_{v_0,w_0} ) \right] \\
& = & \left[p_v (\eta_{v_0,w_1} -  \eta_{v_0,w_0}) + q_v(\eta_{v_1,w_1}  -  \eta_{v_1,w_0})  \right] +  \left({D \over p_wq_w}\right) \left[ p_w(  \eta_{v_1,w_0}  -   \eta_{v_0,w_0})  + q_w(  \eta_{v_1,w_1} -\eta_{v_0,w_1} ) \right].
\end{eqnarray*}

With these results in mind, let us now imagine an idealized population that is identical to the current population in every way, except that there is no LD ($D = 0$) between these loci.   Call the difference in allelic effect sizes ($\beta_v$ and $\beta_w$ in the actual population) $\tilde{\beta_v}$ and $\tilde{\beta_w}$ in the idealized population with no LD.   From the results above we immediately have
\begin{eqnarray*}
\tilde{\beta_v} & = & \left[p_w( \eta_{v_1,w_0}  - \eta_{v_0,w_0} ) + q_w(  \eta_{v_1,w_1} - \eta_{v_0,w_1} \right)].  \\
\tilde{\beta_w} & = &  \left[p_v (\eta_{v_0,w_1} -  \eta_{v_0,w_0}) + q_v(\eta_{v_1,w_1}  -  \eta_{v_1,w_0})  \right]. \\
\beta_v & = & \tilde{\beta_v} + {D \over p_vq_v} \tilde{\beta_w} \\
\beta_w & = & \tilde{\beta_w} + {D \over p_wq_w} \tilde{\beta_v} 
\end{eqnarray*}
In this manner we arrive at the fundamental intuition concerning LD's influence on effect sizes.   The effect size at locus $v$, measured as the difference in average phenotype between individuals with an $A_1$ versus $A_0$ allele at locus $v$, is equal to what the effect size would be at locus $v$, absent LD, plus the effect at locus $w$, absent LD, weighted by the haplotypic covariance between the two loci, divided by the allelic variance at locus $v$, a quantity that might be called the ``LD regression coefficient.''  This is all formally true within our Kempthorne inspired interpretations of allelic effects.   In a more Falconer inspired view, we would likely think of $\tilde{\beta_v}$ and $\tilde{\beta_w}$ as the ``true'' effect sizes at the two loci, with $\beta_v$ and $\beta_w$ being thought of as the ``estimated'' effects in the presence of LD.   With a Falconer view in mind, we might phrase this most simply as the apparent effects at one SNP is the sum of the true affect at the SNP, plus the effects of another SNP times the LD regression coefficient between them.  Whether one thinks of $\tilde{\beta}$ as either the ``true'' effect (in the Falconer sense) or the effect in a population absent LD (in the Kempthorne sense), calculation of $\tilde{\beta}$ could prove extremely useful in applications where effects estimated in one population will be applied to another population with differing LD.    This also suggests a potential approach for accounting for LD in a study.   If we again assume a lack of dominance or interaction from any source other than LD, and further assume that higher order LD is reasonably approximated by pairwise LD, for all SNPs in a given region, we can begin by estimating their effect sizes, in the Kempthorne sense, $\vec{\beta}$.  If we also have estimates of the LD covariance ($D_{v,w}$) between all pairs of sites $[v,w]$, and individual site allele frequencies $p_v,q_v$, we can construct an LD regression matrix $\bold{M}$ with $m_{v,w} = {D_{v,w} \over p_vq_v}$, and use the relationship 
\begin{eqnarray*}
\vec{\beta} & = & \bold{M} \vec{\tilde{\beta}}. \\
\vec{\tilde{\beta}} & = & \bold{M}^{-1} \vec{\beta}. 
\end{eqnarray*} 
\noindent In practice, the LD matrix is likely to be very stiff (frequently with degenerate rows from pairs of sites in perfect LD), so there will necessarily be numerical challenges with implementing this sort of approach, but in principle this idea could be used for explicit accounting for LD, and application of estimates taken from one LD setting into another.

\subsection{Intuition about dominance and interactions}

Dominance is a term used by population geneticists to describe the relationship between the phenotype of the heterozygote and the two homozygotes.   If the heterozygote has a phenotype equal (or nearly equal) to one of the homozygotes, we tend to say the allele associated with the homozygote genotype which is equal to the heterozygote phenotype is ``dominant'' to the other allele.  Conversely we say the allele associated with other homozygote genotype is ``recessive.''  Additivity is a form of partial or incomplete dominance where heterozygote phenotype is between the two homozygous phenotypes.   Over/Underdominance is used to describe heterozygote phenotypes outside the range of the two homozygote (above/below).   

These definitions are well ingrained in population genetics.  Dominance is nearly synonymous with the phenotype of the heterozygote.  As a result there is, perhaps, an intuitive desire to believe that a quantitative locus can be described as either additive, or if not additive with only one additional parameter to describe the heterozygous phenotype, \emph{ala} $1,1-hs,1-s$ in a single locus population genetics scenario.  This is simply not true when the additive effect is defined as the mean phenotype of the allele.  A locus is either additive, in which case all three dominance deviations are $0$, or it is not additive, in which case all $3$ deviations are non-zero.   Any attempt to parameterize this system with only two or fewer values will lead to none of them being interpretable as the additive effect, unless the locus is additive. 

Another important insight is that the size of the dominance variance is very much a function of allele frequency.    The only possible way for the dominance variance to be a large fraction of the total genetic variance is for the rare allele to be significantly recessive, \emph{i.e.} for the heterozygote to have phenotype much closer to the common homozygote phenotype.   This can be intuitive.  Rare alleles are found more often as heterozygotes than homozygotes.  The rarer the truer this is.  So, the mean phenotype of a rare recessive allele tends to be closer to the heterozygote phenotype than the homozygote, which results in greater deviation from additivity.  Intuitively the additive approximation to all $3$ is most in ``error'' when the rare allele is most recessive, and the size of this error increases with increasing rarity of the recessive allele.   Viewed the other way around, for a recessive locus where the recessive allele is common, most of the genetic variance will be additive.  A recessive locus where the recessive allele is rare will have mostly dominance variance.

Finally it should be clear that each of the interaction terms is defined by the difference between the observed mean phenotype and what would be expected under additivity plus all the interactions at a ``higher level''.  Additive by dominance expectations include all the appropriate additive by additive interactions.   Additive by Additive by Additive expectations include all the appropriate additive by additive interactions, \emph{etc}.   Thus, unless something truly perverse is going on with these distributions, it would be natural and expected for each level of interaction to be smaller in magnitude than the previous level.    Effectively each level is the residual variance after accounting for all the main and interaction effects on the previous level.   As a result it is perfectly natural to expect $V_{GGG} < V_{GG} < V_{G}$.     

\section{Application to Human Disease}

Many human ``disease'' phenotypes, diastolic blood pressure, say, is well modeled and understood using the quantitative genetic machinery described above.  Diastolic blood pressure is approximately normally distributed in most studies \cite{bp2011}. Investigators can and frequently do estimate heritability of the trait from family studies (sib-pairs or parents and offspring, say) \cite{hypertension2019} in the manner described above.   At individual SNPs, the effect, $\beta = \alpha_1 - \alpha_0$, of substituting an $A_1$ allele for an $A_0$ is frequently estimated in some sort of regression framework.    If we call this locus $i$, the heritability due to locus $v$, $h^2_v$, can be estimated from this regression analysis \cite{bpgwas2018}].  Recalling as shown above $p\alpha_0 + q\alpha_1 = 0$, 
\begin{eqnarray*}
h^2_v & = & {\mathrm{Var}[a_v] \over V_P}. \\
\mathrm{Var}[a_v] & = & 2(p\alpha_0^2 + q\alpha_1^2) \\
& = & 2 ( p\alpha_0^2(p+q) + q \alpha_1^2(p+q) ) \\
& = & 2( p^2\alpha_0^2 + pq\alpha_0^2 + pq\alpha_1^2 + q^2\alpha_1^2) \\ 
& = & 2( p^2\alpha_0{-q\alpha_1 \over p} + pq\alpha_0^2 + pq\alpha_1^2 + q^2\alpha_1{-p\alpha_0 \over q}) \\ 
& = & 2( -pq\alpha_0\alpha_1 + pq\alpha_0^2 + pq\alpha_1^2 + -pq\alpha_0\alpha_1) \\
& = & 2pq( \alpha_0^2 -2\alpha_0\alpha_1 + \alpha_1^2) \\
& = &  2pq(\alpha_1 - \alpha_0)^2 \\
& = & 2pq\beta^2.
\end{eqnarray*}
Thus, in a standardly designed Genome-Wide Association Study (GWAS) of a quantitative disease phenotype, $P$, such as diastolic blood pressure, is measured in a large number of individuals, and in those same individuals genotype is determined at a large (perhaps $10^6$ or more) number, $n$, of SNPs.   At each locus the $A_0$ and $A_1$ alleles are coded as $0$ and $1$ respectively, and the genotype is coded as the sum of the alleles.   The investigator then performs $n$ independent linear regressions of phenotype as the outcome and genotype as the predictor, including any measured environmental co-variates that correlate with outcome, and often co-variates estimated from the entire genome's genotypes to account for population structure within the study \cite{Price2006}.    Alternatively, and perhaps more technically appropriate, a linear-mixed model might be performed where the rest of the genome's genotype is treated as a random ($\approx V_A$) effect \cite{Lewis2020}.   

The result of this study is $n$ independently estimated $\beta$'s.    If none of these sites were in LD with one another, and no other genetic interactions exist, and there are SNPs in all areas of the genome with genetic contributions to phenotype,  $V_A$, and consequently heritability, could be estimated as $2pq\beta^2$ summed across all SNPs.  This is the insight that lies at the heart of LD Score regression and related methods \cite{Bulik-Sullivan:2015aa}.  Alternatively, $V_A$ could be estimated as the random effect term in a linear mixed model \cite{Parker2016}.  

Somewhat recently, a form of analysis has developed, often called polygenic risk scores (PRS) \cite{Lewis2020} or some related phrase, that is frequently useful.  In this form of analysis, $\beta$'s are usually estimated in one study, and then in a second study, individuals with known genotype have their expected phenotype calculated using the first study's $\beta$'s.    Details and challenges associated with this style of analysis will be discussed in much greater detail in the second in this series of papers.

\subsection{Binary Traits}

In many ways, the field of human genetics arose largely independently of any quantitative genetics ideas.   For much of its early history \cite{Harris1963,McKusick:1998aa}  the field was largely concerned with understanding nearly binary traits (traits with only $2$ major phenotypes) under nearly Mendelian control (single locus genetics).  At first glance, there was no obvious connection between the modeling framework presented here, which results in approximately normally distributed phenotypes, and the approximately binary traits that were of deepest interest to human geneticists. 

In 1965 in a seminal work by Falconer \cite{falconer1965}, the natural connection between human binary phenotypes and the importance of quantitative genetics to understanding them was first presented in detail.   The key idea was to suppose that a binary phenotype is like any other quantitative phenotype, but observed on ``the wrong scale.''   For any binary trait of interest, Crohns Disease (CD), say, humans are characterized as either having CD, or not.  However, following Falconer, a quantitative geneticist will think about CD like any other quantitative trait.   To do so, they will assume there is a related trait which they will generally call ``liability'' to CD.   This trait, liability to CD, is a quantitative trait like any other.  It is contributed to by genes and the environment.  Its variance components can be decomposed as described above.   However, liability is not directly observable.  We do not observe or measure liability to CD directly.  Instead we observe the effects of the existence of a threshold $t$ (FIGURE 2).  Individuals with liability greater than or equal to $t$ we observe to have CD.   Individuals with liability less than $t$ we say do not have CD.  

In our personal experience, many physician scientists will immediately express skepticism about the applicability or utility of this abstraction, ``liability to disease,'' to their particular areas of study.  Interestingly, one of the first implications of this abstraction is that there ought to exist individuals with liability very near the threshold.   Presumably such individuals will often be very hard to classify.  They are ``unaffected'' people who nearly have the disease, or they are affected people who have only a very mild form of the disease.  These are individuals who two well trained physicians might reasonably disagree on whether or not such a person formally qualifies for diagnosis of the disease.   Viewed in this light, we can see the abstraction of an unobservable liability is the cause for the existence of individuals who either slightly do, or do not, reach diagnostic criteria for a disease.   Such individuals have liability very near the threshold, and because liability is unobservable directly, two perfectly well trained physicians may disagree about which side of the threshold a particular individual lies.  

Because liability is unobserved directly, it is extremely useful to impose assumptions on its distribution that make our modeling of it easier.  First and foremost, we assume the distribution of liability in the population is exactly a normal distribution (or sometimes is exactly normally distributed other than from the effects of a single factor under consideration \cite{morton1974}.  While it is certain that many (most) observable traits are nearly normally distributed \cite{Barton:1989aa} , the assumption of complete convergence in distribution to normality is a far stronger assumption than we have made up to this point.  That meaningful departure from this assumption may not be particularly\cite{Turelli:1994aa} common is reassuring.  Thus, here for the first time we assume a fully normally distributed trait, which we call liability to some binary phenotype, often a human disease.   Because this normally distributed trait is unobserved, we can assume it is parameterized in anyway we please.    For convenience we will assume that liability has mean $0$, and total variance $V_P = 1$, \emph{i.e.} follows a ``standard'' normal distribution.     For such traits heritability $h^2 = V_A/V_P = V_A$.   Thus, it will not be uncommon for human quantitative geneticists to call something heritability or a contribution to heritability, while clearly estimating $V_A$, or $V_{a_v}$ the additive variance due to locus $v$.   In fact, as a field $V_{a_v}$ is often called the ``SNP heritability'' of locus $v$.   When $V_P = 1$ heritability and additive variance are identical in value, but interchanging their terms can certainly lead to confusion, particularly of students new to the field.  Finally, for convenience we will always orient the threshold $t$ to be a positive value.   Thus, by convention increasing liability increases the chance of disease, and decreasing liability has the opposite effect.
 
\subsection{Human Disease Quantities}

The human genetics field often has its own set of terms of art that are sometimes confusing to classically trained population or quantitative geneticists.   Above we saw that human geneticists often call Wright's IBD probabilities Cotterman coefficients.   Here, for the sake of explicit understanding, we will define several terms that frequently occur in human disease studies.  

We begin by assuming there is a population of humans that at least approximately corresponds to a single, finite Fisher-Wright population in Hardy-Weinberg equilibrium.  In this population, there is a quantitative phenotype $L$, which is the liability to some disease of interest.   There is a threshold, $t$, on this liability scale such that individuals with $L \ge t$ are said to be diseased, and individuals with liability below $t$ are said to be ``healthy'' or not to have the disease in question. The term ``prevalence'' of a disease, $\psi$ is the fraction of the population with disease and is uniquely determined by $t$, 
\begin{eqnarray*}
\psi & = & \int_{t}^{\infty} \phi(x) dx \\
& = & 1 - \Phi(t) \\
t & = & \Phi^{-1}(1-\psi),
\end{eqnarray*}
\noindent where $\phi(x)$ is a standard normal probability density, $\Phi(x)$ is a standard normal cumulative distribution, and $\Phi^{-1}(x)$ is its inverse.   Thus, we think of the prevalence of a disease as determining the threshold on the liability scale beyond which individuals are diseased.

One of the key questions in human genetics is ``What effect does a given SNP have on disease liability?''  Within our Kempthorne framework, we imagine this effect causes the mean liability of individuals with different genotypes to differ  [FIGURE 3].   If we could observe liability directly, we could immediately apply all of the previous machinery.   Here, though, liability is not directly observed.   Instead, in the classical human genetics experiment, a number $N_D$ people with disease are identified along with $N_{\not D}$ people without the disease.  By convention people with the disease are often called ``cases'' and people without the disease called ``controls.'' Cases and controls are often collected in a very biased way relative to disease prevalence.   Usually cases are dramatically oversampled such that $N_D \gg \psi N_{\not D}$.   Regardless of the sampling proportions, the fundamental data collected is the counts $n_{00}$,$n_{01}$,$n_{11}$ of the three genotypes $A_0A_0$,  $A_0A_1$,  $A_1A_1$, broken down by case, $n_{ij}^D$, and control $n_{ij}^{\not D}$, $n_{ij} =  n_{ij}^D + n_{ij}^{\not D}$ status.  It is perhaps not immediately intuitive, but given disease prevalence $\psi$, these counts  are sufficient to estimate all of the above described quantitative genetics quantities.  

Start by calculating genotype $f_{ij}$ and allele $p,q$ frequencies within case and control subsets, and from those estimating the total population quantities.  
\begin{eqnarray*}
f_{ij}^D & = & {n_{ij}^{D} \over N_D} \\
f_{ij}^{\not D} & = & {n_{ij}^{\not D} \over N_{\not D}} \\
p^{D} & = & f_{00}^D + {f_{01}^D \over 2} \\ 
q^{D} & = & 1 - p^D \\
p^{\not D} & = & f_{00}^{\not D} + {f_{01}^{\not D} \over 2} \\ 
q^{\not D} & = & 1 - p^{\not D} \\
f_{ij} & = & \psi f_{ij}^D  + (1 - \psi)  f_{ij}^{\not D} \\
p & = & \psi p^{D} + (1-\psi) p^{\not D} \\
q & =&  1- p.
\end{eqnarray*}
The term penetrance of $X$ is the conditional probability of an individual being diseased given they are in state $X$.  Thus, we can consider the penetrance $\zeta$ of a genotype $G_{ij}$, the probability an individual is diseased given their genotype is $A_iA_j$ at this locus.  We can also think about penetrance of an allele $A_i$, the probability an individual is diseased given they have an $A_i$ allele.   Thus,
\begin{eqnarray*}
\zeta_{G_{ij}} & = & \mathrm{Pr}[L > t | G = A_iA_j] \\
& = & \mathrm{Pr}[D | G = A_iA_j ] \\
\zeta_{A_i} & = &  \mathrm{Pr}[L > t | A = A_i] \\
& = & \mathrm{Pr}[D | A = A_i ]. 
\end{eqnarray*}
\noindent With application of Bayes' theorem, penetrances can be immediately estimated from the case/control data. 
\begin{eqnarray*}
\mathrm{Pr}[D \cap G = A_iA_j] & = & \mathrm{Pr}[G =  A_iA_j | D] \mathrm{Pr}[D] \\ 
& = & \mathrm{Pr}[D | G =  A_iA_j] \mathrm{Pr}[ G =  A_iA_j]. \\
\zeta_{G_{ij}} & = & {f_{ij}^D \psi \over f_{ij}} \\
\zeta_{A_0} & = & {p^{D} \psi \over p } \\
\zeta_{A_1} & = & {q^{D} \psi \over q }
\end{eqnarray*} 
\noindent Thus, from the overall prevalence and genotype counts in cases and controls, we can estimate the penetrance (probability of disease given genotype/allele) of both alleles, and all three genotypes.   Of course, as quantitative geneticists we measure effect sizes in terms of mean effects on \emph{liability}, but that too in now immediately available, with a sensible approximation, or can be found numerically.   To find this, recall that we have normalized liability to have $V_P = 1$.  If the three genotype at this locus have mean liability $\gamma_{00}$, $\gamma_{01}$, and $\gamma_{11}$ respectively, then 
\begin{eqnarray*}
V_g & = & f_{00} \gamma_{00}^2 + f_{10} \gamma_{01}^2 + f_{11} \gamma_{11}^2. \\
\zeta_{G_{ij} } & = & \int_{t}^{\infty} \phi(\gamma_{ij},1-V_g) dx \\
& = & \int_{t-\gamma_{ij}}^{\infty} \phi(0,1-V_g) dx \\
& \approx & \int_{t-\gamma_{ij}}^{\infty} \phi(0,1) dx. \\
\gamma_{ij} & \approx & t - \Phi^{-1}(1-\zeta_{G_{ij} }). \\
\alpha_{i} & \approx &  t - \Phi^{-1}(1-\zeta_{A_{i} }),
\end{eqnarray*}
where $\phi(\mu,\sigma^2)$ is a normal density with mean $\mu$ and variance $\sigma^2$.  The above approximations hold whenever $V_g \ll 1$.   Since for the vast majority of human disease \cite{Loos:2020aa}  there are few sites that explain even $0.1\%$ of the variance, this approximation is almost always very good.   When trying to estimate something that explains a truly substantial fraction of the variance, a Newton-Raphson iteration (or just about any other kind of numerical search) will converge quickly.  Nevertheless, even for very small genetic variances it is often useful to estimate ``all but one'' of the mean effects, and find the remaining effect using the fact that the average effect must be zero.  Thus, it is often helpful to estimate these effects as
\begin{eqnarray*}
\gamma_{11} & = & t - \Phi^{-1}(1-\zeta_{G_{11} })). \\
\gamma_{01} & = & t - \Phi^{-1}(1-\zeta_{G_{01} })). \\
\gamma_{00} & = & { -(f_{11} \gamma_{11} + f_{01}\gamma_{01}) \over f_{00}}. \\
\alpha_1 & = &   t - \Phi^{-1}(1-\zeta_{A_1 })). \\
\alpha_0 & = & {-q\alpha_1 \over p}
\end{eqnarray*}

Calculating effects in this manner assures that the population mean remains $0$ despite the approximation used for the residual variance.    Thus, starting with only prevalence and the counts of genotypes we have arrived at all the quantitative genetic quantities needed to calculate additive and dominance contributions to variance.   Higher order interactions can be approached the same way, via counts of individuals with two (or more) locus genotypes, divided between cases and controls.   

Historically effect sizes in human genetics tend to be reported as either a ``relative risk.'' or an ``odds ratio.''  Both quantitates are some sort ratio of the penetrances.    In general, the relative risk of X to Y, is ${\mathrm{Pr}[D | X] \over \mathrm{Pr}[D | Y]}$, \emph{i.e.} it is the ratio of the penetrance of X to the penetrance of Y.   Building on historical gambling terms, the ``odds'' of something is the probability the event happens, divided by the probability the event does not happens.   Thus the odds of X are ${\mathrm{Pr}[D | X] \over 1-\mathrm{Pr}[D | X]}$.   So, the odds ratio of X to Y is ${\mathrm{Pr}[D | X] (1- \mathrm{Pr}[D|Y]) \over \mathrm{Pr}[D | Y](1-\mathrm{Pr}[D | X])}$.  Thus, it might be natural to discuss the odds ratio of the $A_1$ allele to the $A_0$ allele, or even the $G_{11}$ genotype to the $G_{00}$ genotype, say.  

For very practical reasons the odds ratio of $A_1$ to $A_0$ (or the other way around) is the most commonly reported effect size estimate in all human genetics studies.  The reason for this is that odds ratios ($\mathrm{OR}$) can be estimated in the presence of covariates in a very natural way.  Recall for a classically observed quantitative phenotype we might commonly estimate $\beta$ for a SNP from a linear regression (or even linear mixed model) that included any covariates known to correlate with phenotype, such as some measured environmental variable (or related quantity such as sex or age), and almost always including estimates of genome-wide genotype to account for population structure (the fact that not all samples come from a single idealized randomly mating population).   The outcome of this linear regression is an estimate of the mean effect $\beta$ of substituting an $A_1$ allele for an $A_0$ allele on phenotype.  From strictly genotype count data it is hard to immediately imagine a framework that would allow estimation of $\beta$ while accounting for covariates in a similar fashion.  Logistic regression turns out to be the non-obvious, but extremely practically useful approach to the problem.

To understand why, FIGURE 4 plots the penetrance on the Y-axis of an allele with mean liability ($Z$, shown in standard deviations where a standard deviation is ${4 \over \sqrt{2\pi}})$ on the X-axis for a trait with prevalence $0.5$ and threshold $t = 0.0$ versus a standard logistic curve (${1 \over 1 + e^{-x}}$).   While this is for a very specific normal distribution, the intuition we form from this is that if liability is well approximated by a normal distribution then the penetrance for an allele is likely well approximated by a logistic function.    Logistic regression is a relatively simple and widely available numerical procedure to estimate the odds ratio of $A_1$ to $A_0$ from case/control count data by fitting the observations to logistic curves for the penetrances of each allele.  This is done without having to know prevalence, or even overall allele frequency, and the estimate can account for the effects of any number of covariates as simply as ordinary linear regression.     The practicality and the utility of this approach should be clear.

To a quantitative geneticist the output of a logistic regression (the odds ratio $\mathrm{OR}$) is not particularly useful \emph{per se}.   Absent knowledge of the disease prevalence, it can only be viewed as an approximation to an interesting but not particularly interpretable quantity.   However, if disease prevalence is known (or estimated) the odds ratio can be converted into our standard measures of effect.   To do so we note that 
\begin{eqnarray*}
\mathrm{Pr}[D] & = & \mathrm{Pr}[D | A_0] \mathrm{Pr}[A_0] +  \mathrm{Pr}[D | A_1] \mathrm{Pr}[A_1]. \\
\psi & = &p \zeta_{A_0} + q  \zeta_{A_1}. \\
\zeta_{A_0} & = & {\psi - q \zeta_{A_1}  \over 1-q}. \\
\mathrm{OR} & = & { \zeta_{A_1} (1-\zeta_{A_0}) \over \zeta_{A_0} (1- \zeta_{A_1})}.\\
\end{eqnarray*}
\noindent From the above one can solve for $\zeta_{A_0}$, albeit in a painful blizzard of algebra involving quadratic terms.    Usually one assumes that the common allele has a penetrance nearly equal to population prevalence and reaches
\begin{eqnarray*}
\mathrm{OR} & \approx & {\zeta_{A_1} (1 - \psi) \over \psi(1-\zeta_{A_1}) } \\
\zeta_{A_1} & \approx & {\mathrm{OR} \psi \over 1 + \psi(\mathrm{OR}-1)} \\ 
\zeta_{A_0} & = &  {\psi - q \zeta_{A_1}  \over 1-q}. \\
\end{eqnarray*}
\noindent Of course, one could numerically iterate these $\zeta$'s to converge to the exactly estimated $\mathrm{OR}$, but given that the logistic curve  itself is an approximation to penetrance of a normally distributed liability, seeking such precision seems a bit like overkill.  Estimated in this fashion the two allelic penetrances are consistent with the overall prevalence of the disease, and for anything other than absurdly large effect sizes, have odds ratio close to the estimated value from the logistic regression.  With the estimates of penetrances in hand, we can convert back to mean effects on the liability scale, and again use all of our standard quantitative genetics ideas to arrive at notions such as SNP heritability \emph{etc} estimated from a logistic regression with case/control counts.

\subsection{Heritability of a disease}

As first discussed by Falconer \cite{falconer1965} , this same framework allows us to estimate overall heritability of any binary phenotype such as a human disease.  To do so, one first needs an estimate of disease prevalence $\psi$, and the disease threshold $t$, found as described above.   Interestingly, and perhaps not instantly obvious, the disease threshold allows calculation of the average liability, $E[L | D]$, of affected individuals.
\begin{eqnarray*}
\mathrm{E}[L | D] & = &  \int_{t}^{\infty} x \phi(x) dx \\
& = & {\phi(t) \over \psi}.
\end{eqnarray*}  
\noindent Thus, the mean liability of affected individuals is determined by the prevalence of disease.  To this one adds data on affected pairs of individuals with a known familial relationship, for instance, pairs of siblings both affected with the disease, or a parent and offspring both affected, \emph{etc}.  The basic design is to first identify individuals with the disease. Such an individual is often called the ``proband.''  Identification of probands, being predicated on disease state, is necessarily biased relative to overall disease prevalence, but is assumed to be an unbiased collection of diseased individuals.  Thus, probands are assumed to have average liability, $\mathrm{E}[L|D]$, as given above.   Once identified, relatives of specific relatedness $\rho$ to the proband are then identified as completely as possible, and the affectation status of these relatives is ascertained.  For instance, the relatives might be a parent of the proband such that $\rho=0.5$.   The faction of these relatives $\zeta_{\text{relative}}$ who are also affected with disease is estimated.   This fraction, $\zeta_{\text{relative}}$, is an estimate of the penetrance of disease given the individual is the specified degree of relatedness to the proband.  Thus, $\zeta_{\text{relative}} = \mathrm{Pr}[D | \text{relative}]$, and we can find the mean liability of these relatives $\mathrm{E}[L | \text{relative}]$ with
\begin{equation*}
\mathrm{E}[L | \text{relative}] =  t - \Phi^{-1}(1-\zeta_{\text{relative}}). 
\end{equation*}
\noindent In this manner we now have the mean phenotype of pairs of relatives with known relatedness $\rho$.   We can then estimate disease heritability $h^2$ in the ``usual'' manner,
\begin{equation*}
h^2 = {\mathrm{E}[L | \text{relative}]  \over \rho \mathrm{E}[L | D]}.
\end{equation*}
  
\section{Acknowledgments}

Population geneticists of a certain age will, without doubt, recognize the voice and influence of John H. Gillespie throughout this presentation.   While attributed here to Kempthorne, because that is how he attributed it, it is certain that virtually everything in the Definitions and Foundational Results section has at least transiently appeared on a blackboard, 12 inches in front of a chalk covered Michael Turelli, feverishly deriving the next result with his right hand while erasing older results with his left.  This work has been supported by NIH Grant RF1 AG071170. 

\newpage
\bibliographystyle{apalike}
\bibliography{qghd}

\begin{landscape}
\thispagestyle{plain}
\begin{center}
\begin{table}
\caption {Summary of key variables} 
\begin{tabular}{ c c c }
Symbol & Description & Formalism \\ \hline 
$A_{v_0}$ & The major allele at locus $v$ & $\mathrm{Freq}[A_{v_0}] \ge  \mathrm{Freq}[A_{v_1}]$ \\ [5pt]
$A_{v_1}$ & The minor allele at locus $v$ & $\mathrm{Freq}[A_{v_1}] \le  \mathrm{Freq}[A_{v_0}]$ \\ [5pt]
$p_v$ &  Frequency of  $A_{v_0}$  & $p_v = \mathrm{Freq}[A_{v_0}]$ \\ [5pt]
$q_v$ & Frequency of  $A_{v_1}$   & $q_v = 1- p_v$ \\ [5pt]
$A_v$ & Random allele at locus $v$, & \\ 
&  $A_v \in \{A_{v_0},A_{v_1}\}$ & $\mathrm{Pr}[A_v = A_{v_0}] = p_v, \mathrm{Pr}[A_v = A_{v_1}] = q_v$ \\ [5pt]
$G_v$ & Random genotype at locus $v$, & $\mathrm{Pr}[G_v = A_{v_0}A_{v_0}] = p_v^2,  \mathrm{Pr}[G_v = A_{v_0}A_{v_1}] = 2p_vq_v$ \\ 
& $G_v \in \{ A_{v_0}A_{v_0},  A_{v_0}A_{v_1}, A_{v_1}A_{v_1} \} $ & $\mathrm{Pr}[G_v = A_{v_1}A_{v1}] = q_v^2$ \\ [5pt]
$\gamma_{v_{00}}$ & Genotypic effect of $A_{v_0}A_{v_0}$ & $\gamma_{v_{00}} = E[P | G_v = A_{v_0}A_{v_0}]$  \\ [5pt]
$\gamma_{v_{01}}$ & Genotypic effect of $A_{v_0}A_{v_1}$ & $\gamma_{v_{01}} = E[P | G_v = A_{v_0}A_{v_1}]$ \\ [5pt]
$\gamma_{v_{11}}$ & Genotypic effect of $A_{v_1}A_{v_1}$ & $\gamma_{v_{11}} = E[P | G_v = A_{v_1}A_{v_1}]$  \\ [5pt]
$\alpha_{v_0}$ & Allelic effect of $A_{v_0}$ &  $\alpha_{v_0} = E[P | A_v = A_{v_0}]$ \\ [5pt]
$\alpha_{v_1}$ & Allelic effect of $A_{v_1}$ &  $\alpha_{v_1} = E[P | A_v = A_{v_1}] = {-p_v\alpha_{v_0} \over q_v}$ \\ [5pt]
$\beta_v$ & Difference in allelic effects & $\beta_v = \alpha_{v_1} - \alpha_{v_0}$ \\ [5pt]
$\delta_{v_{00}}$ & Dominance deviation of genotype $A_{v_0}A_{v_0}$ & $\delta_{v_{00}} = \gamma_{v_{00}} - 2\alpha_0$ \\ [5pt]
$\delta_{v_{01}}$ & Dominance deviation of genotype $A_{v_0}A_{v_1}$ & $\delta_{v_{01}} = \gamma_{v_{01}} - (\alpha_0 + \alpha_1)$ \\ [5pt]
$\delta_{v_{11}}$ & Dominance deviation of genotype $A_{v_1}A_{v_1}$ & $\delta_{v_{11}} = \gamma_{v_{11}} - 2\alpha_1$ \\ [5pt]
$g_v$ & Random genetic effect determined by $G_v$ & If $G_v = A_{v_i}A_{v_j}$, then $g_v = \gamma_{v_{ij}}$ \\ [5pt]
$a_v$ & Random additive effect determined by $G_v$ & If $G_v = A_{v_i}A_{v_j}$, then $a_v = \alpha_{v_i} + \alpha_{v_j}$ \\ [5pt]
$d_v$ & Random dominance deviation determined by $G_v$  & If $G_v = A_{v_i}A_{v_j}$, then $d_v = \delta_{v_{ij}}$ \\ [5pt]
$V_{g_v}$ & Total genetic variance of locus $v$ & $V_{g_v} = p_v^2(\gamma_{v_{00}})^2 + 2p_vq_v(\gamma_{v_{01}})^2 + q_v^2(\gamma_{v_{11}})^2$ \\ [5pt]
$V_{a_v}$ & Additive variance of locus $v$ & $V_{a_v} = 2(p_v\alpha_{v_0}^2 + q_v\alpha_{v_1}^2)$ \\ [5pt]
$V_{d_v}$ & Dominance variance of locus $v$  & $V_{d_v} = p_v^2(\delta_{v_{00}})^2 + 2p_vq_v(\delta_{v_{01}})^2 + q_v^2(\delta_{v_{11}})^2$
\end{tabular}
\end{table}

\thispagestyle{plain}
\begin{table}
\caption*{Table 1: Continued} 
\begin{tabular}{ c c c }
Symbol & Description & Formalism \\ \hline 
$\gamma_{v_{00},w_{00}}$ & Genotypic effect of $A_{v_0}A_{v_0}$ and $A_{w_0}A_{w_0}$  & $\gamma_{v_{00},w_{00}} = E[P | G_v = A_{v_0}A_{v_0}, G_w = A_{w_0}A_{w_0}]$  \\ [5pt]
$\gamma_{v_{00},w_{01}}$ & Genotypic effect of $A_{v_0}A_{v_0}$ and $A_{w_0}A_{w_1}$  & $\gamma_{v_{00},w_{01}} = E[P | G_v = A_{v_0}A_{v_0}, G_w = A_{w_0}A_{w_1}]$  \\ [5pt]
$\gamma_{v_{00},w_{11}}$ & Genotypic effect of $A_{v_0}A_{v_0}$ and $A_{w_1}A_{w_1}$  & $\gamma_{v_{00},w_{11}} = E[P | G_v = A_{v_0}A_{v_0}, G_w = A_{w_1}A_{w_1}]$  \\ [5pt]
$\gamma_{v_{01},w_{00}}$ & Genotypic effect of $A_{v_0}A_{v_0}$ and $A_{w_0}A_{w_0}$  & $\gamma_{v_{01},w_{00}} = E[P | G_v = A_{v_0}A_{v_1}, G_w = A_{w_0}A_{w_0}]$  \\ [5pt]
$\gamma_{v_{01},w_{01}}$ & Genotypic effect of $A_{v_0}A_{v_1}$ and $A_{w_0}A_{w_1}$  & $\gamma_{v_{01},w_{01}} = E[P | G_v = A_{v_0}A_{v_1}, G_w = A_{w_0}A_{w_1}]$  \\ [5pt]
$\gamma_{v_{01},w_{11}}$ & Genotypic effect of $A_{v_0}A_{v_1}$ and $A_{w_1}A_{w_1}$  & $\gamma_{v_{01},w_{11}} = E[P | G_v = A_{v_0}A_{v_1}, G_w = A_{w_1}A_{w_1}]$  \\ [5pt]
$\gamma_{v_{11},w_{00}}$ & Genotypic effect of $A_{v_1}A_{v_1}$ and $A_{w_0}A_{w_0}$  & $\gamma_{v_{11},w_{00}} = E[P | G_v = A_{v_1}A_{v_1}, G_w = A_{w_0}A_{w_0}]$  \\ [5pt]
$\gamma_{v_{11},w_{01}}$ & Genotypic effect of $A_{v_1}A_{v_1}$ and $A_{w_0}A_{w_1}$  & $\gamma_{v_{11},w_{01}} = E[P | G_v = A_{v_1}A_{v_1}, G_w = A_{w_0}A_{w_1}]$  \\ [5pt]
$\gamma_{v_{11},w_{11}}$ & Genotypic effect of $A_{v_1}A_{v_1}$ and $A_{w_1}A_{w_1}$  & $\gamma_{v_{11},w_{11}} = E[P | G_v = A_{v_1}A_{v_1}, G_w = A_{w_1}A_{w_1}]$  \\ [5pt]
$g_{v,w}$ & Random two locus genetic effect  & If $G_v = A_{v_i}A_{v_j},G_w = A_{w_k}A_{w_l}$ \\ 
& determined by genotypes $G_v$ and $G_w$ &  then $g_{v,w} = \gamma_{{v_{ij}},{w_{kl}}}$ \\[5pt]
$\delta_{Ig_{v_{ij},w_{kl}}}$ & Epistatic Deviation & $\delta_{Ig_{v_{ij},w_{kl}}} =  \gamma_{{v_{ij}},{w_{kl}}} - (\gamma_{v_{ij}} + \gamma_{w_{kl}})$ \\[5pt]
$\delta_{Iaa_{v_i,w_k}}$ & Additive by Additive Deviation & $\delta_{Iaa_{v_i,w_k}} = E[P | A_v = A_{v_i}, A_w=A_{w_k}] -  (\alpha_{v_i} + \alpha_{w_k})$ \\[5pt]
$\delta_{Iad_{v_i,w_{kl}}}$ & Additive by Dominance Deviation & $\delta_{Iad_{v_i,w_{kl}}} = E[P | A_v = A_{v_i}, G_w=A_{w_k}A_{w_l}]$ \\ 
& & $-(\alpha_{v_i} + \alpha_{w_k} + \alpha_{w_l} + \delta_{w_{kl}})$  \\[5pt]
$\delta_{Idd_{v_{ij},w_{kl}}}$ & Dominance by Dominance Deviation& $\delta_{Idd_{v_{ij},w_{kl}}} =  E[P | G_v = A_{v_i}A_{v_j}, G_w=A_{w_k}A_{w_l}]$ \\
& & $- (\alpha_{v_i} + \alpha_{v_j} + \delta_{v_{ij}} + \alpha_{w_k} + \alpha_{w_l} + \delta_{w_{kl}})$ \\ [5pt]
$L$ & An unobserved phenotype, liability to disease & $L \sim \Phi(x)$ \\[5pt]
$t$ & A threshold on the liability & Individual is diseased if $L \ge t$ \\
&  scale determining disease & \\[5pt]
$\psi$ & Prevalence of the disease with liability $L$ & $\psi = \int_{t}^{\infty} \phi(x) dx$ \\[5pt]
$\zeta_y$ & The penetrance of some factor $y$ & $\zeta_y = Pr[L \ge t | y]$ \\[5pt]
\end{tabular}
\end{table}
\end{center}
\end{landscape}

\begin{figure} 
\includegraphics[max size={12in}{\textheight}]{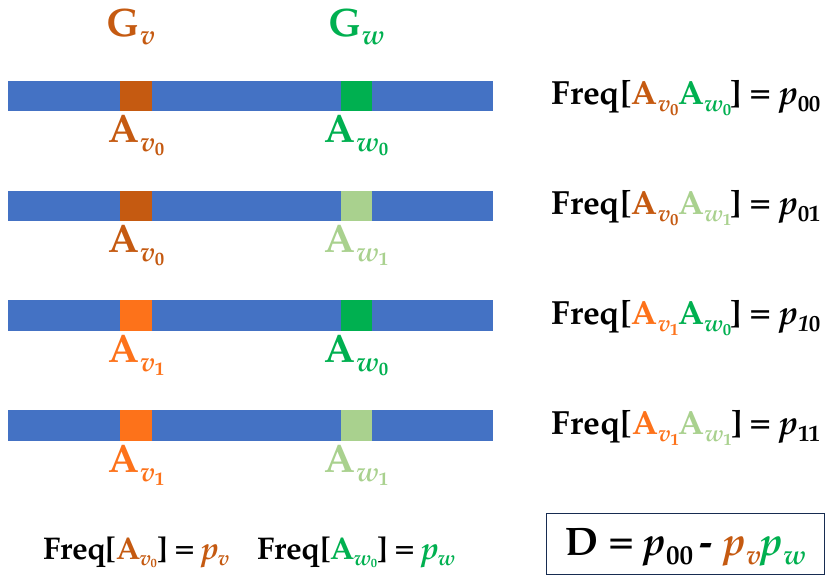}
\caption{Two locus LD.} 
\end{figure}

\begin{figure} 
\includegraphics[max size={7in}{\textheight}]{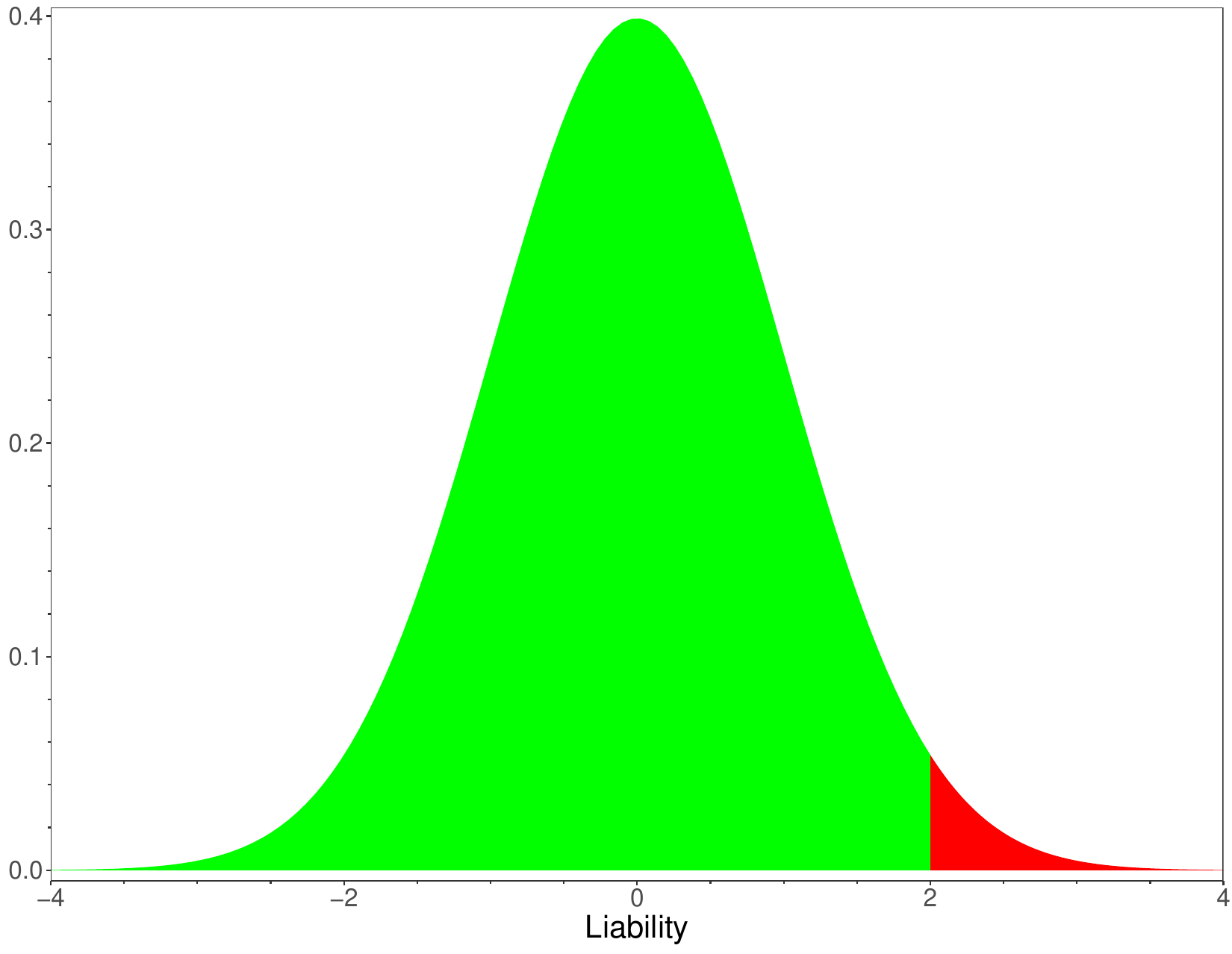}
\caption{Normally distributed liability with disease determining threshold at liability greater than or equal to $2$.} 
\end{figure}

\begin{figure} 
\includegraphics[max size={8in}{\textheight}]{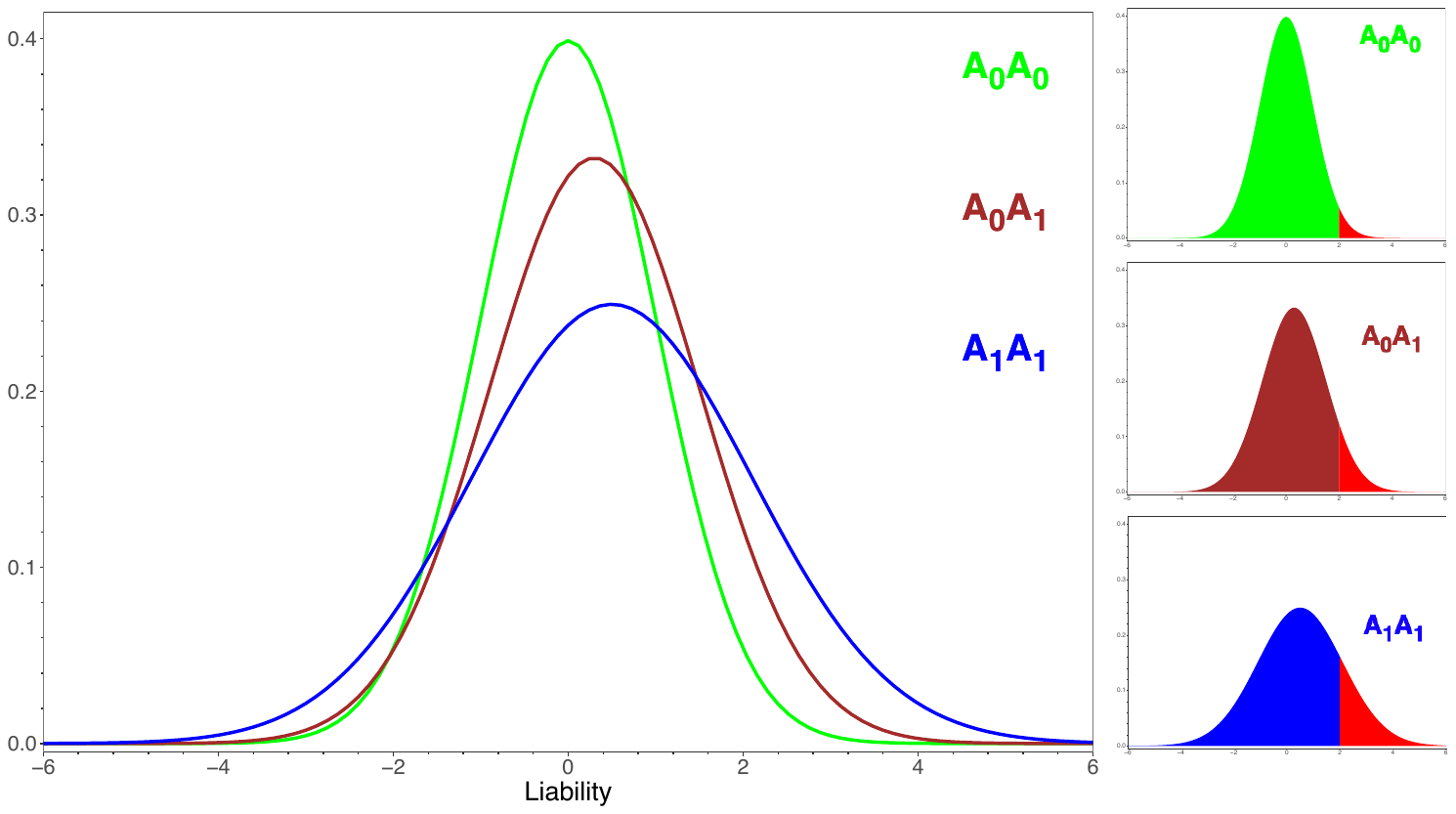}
\caption{Genotypes with differing mean liability have differing penetrances.} 
\end{figure}

\begin{figure} 
\includegraphics[max size={6in}{\textheight}]{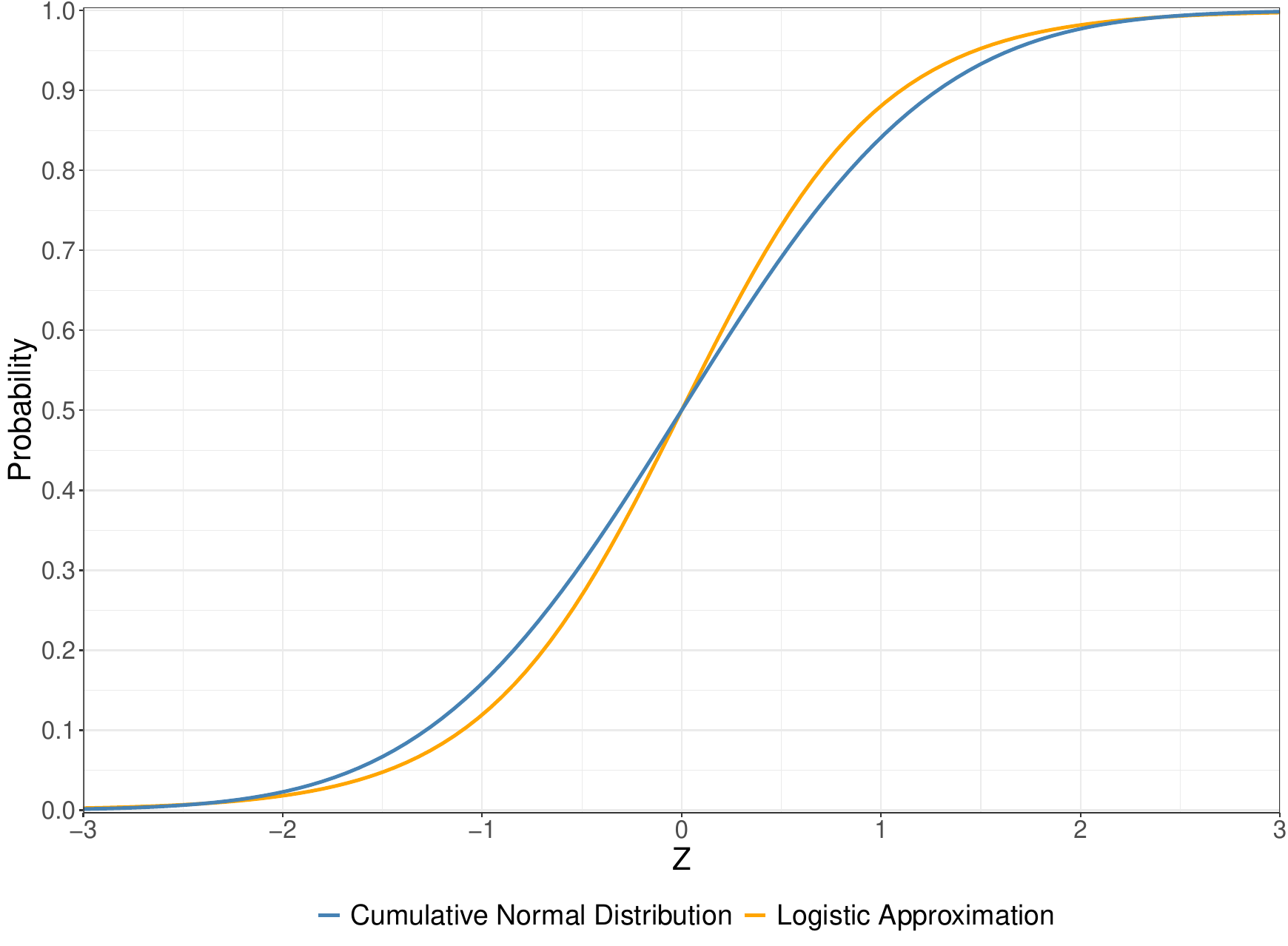}
\caption{Penetrance assuming normally distributed liability versus logistic approximation for a trait with threshold at $0$.} 
\end{figure}

\end{document}